\documentclass[%
 reprint,
 superscriptaddress,
 graphicx,
 amsmath,amssymb,
 aps,prr,
 longbibliography,
]{revtex4-2}

\usepackage{graphicx}% Include figure files
\usepackage{dcolumn}% Align table columns on decimal point
\usepackage{bm}% bold math
\usepackage[breaklinks=true,colorlinks=true,linkcolor=blue,urlcolor=blue,citecolor=blue]{hyperref}
\usepackage{upgreek}

\begin{document}

%\preprint{APS/123-QED}

% \title{Embedding fundamental physical symmetries into data-driven reduced plasma models via data augmentation}
\title{Embedding physical symmetries into machine-learned reduced plasma physics models via data augmentation}
%alternative titles
% Generation of high-quality ion beams from radiation pressure acceleration of thin targets
% High-quality radiation pressure ion acceleration from thin targets via the mitigation of electron heating
\author{M. C. McGrae-Menge}
\email[]{madoxmm@physics.ucla.edu} 
\affiliation{Department of Physics and Astronomy, University of California, Los Angeles, CA 90095, USA}
\author{J. R. Pierce}
\affiliation{Department of Physics and Astronomy, University of California, Los Angeles, CA 90095, USA}
\author{F. Fiuza}
\affiliation{
 GoLP/Instituto de Plasmas e Fus\~ao Nuclear, Instituto Superior T\'ecnico, Universidade de Lisboa, 1049-001 Lisbon, Portugal
}
\author{E. P. Alves}
\affiliation{Department of Physics and Astronomy, University of California, Los Angeles, CA 90095, USA}
\affiliation{Mani L. Bhaumik Institute for Theoretical Physics, University of California at Los Angeles, Los Angeles, CA 90095, USA}

\date{June 18, 2025}% It is always \today, today,

\begin{abstract}
Machine learning is offering powerful new tools for the development and discovery of reduced models of nonlinear, multiscale plasma dynamics from the data of first-principles kinetic simulations. However, ensuring the physical consistency of such models requires embedding fundamental symmetries of plasma dynamics. In this work, we explore a symmetry-embedding strategy based on data augmentation, where symmetry-preserving transformations (e.g., Lorentz and Galilean boosts) are applied to simulation data. Using both sparse regression and neural networks, we show that models trained on symmetry-augmented data more accurately infer the plasma fluid equations and pressure tensor closures from fully kinetic particle-in-cell simulations of magnetic reconnection. We show that this approach suppresses spurious inertial-frame-dependent correlations between dynamical variables, improves data efficiency, and significantly outperforms models trained without symmetry-augmented data, as well as commonly used theoretical pressure closure models. Our results establish symmetry-based data augmentation as a broadly applicable method for incorporating physical structure into machine-learned reduced plasma models.
\end{abstract}

\maketitle

%\tableofcontents

\section{\label{sec:intro}Introduction}
Plasmas are hot, ionized gasses that exhibit complex nonlinear behavior across a broad range of spatial and temporal scales. Notable examples include inertial and magnetically confined fusion plasmas~\cite{Nuckolls1972,Ongena2016}, where the coupling between microturbulence and transport scales plays a central role in plasma confinement, and astrophysical plasmas where dissipation of large-scale magnetic fields is regulated by small kinetic-scale plasma processes that further mediate the acceleration of energetic particles ~\cite{Sironi2014,Guo2015}. A comprehensive understanding of such multiscale plasma phenomena remains a major scientific challenge, requiring improved theoretical frameworks and modeling tools to make progress. First-principles approaches based on plasma kinetic theory and simulations offer accurate descriptions of plasma dynamics but are computationally prohibitive for capturing the full range of scales in many realistic systems of interest. In contrast, fluid models are efficient at capturing large-scale plasma behavior but often miss important effects associated with small-scale kinetic processes.  To effectively model such multiscale plasma dynamics it is important to develop reduced plasma models that capture the key coupling between small-scale kinetic processes and large-scale plasma fluid behavior, without explicitly resolving the small scales.

Traditionally, reduced plasma models have been developed using analytic theory, which are typically valid in linear plasma dynamics regimes \cite{Hammett1990} or in well-defined asymptotic limits \cite{Ohia2012}. However, the assumptions that underpin these models often break down in many scenarios of interest, where the physics is nonlinear and multiscale. Current state-of-the-art kinetic plasma simulations are beginning to be able to capture sufficiently large dynamical ranges to study nonlinear multiscale plasma dynamics from first principles \cite{Comisso2018,Alves2018}. The data from these simulations is creating new opportunities to guide the development of reduced plasma descriptions in these regimes.

Machine learning (ML) techniques are emerging as promising tools to develop reduced plasma models from the data of first-principles simulations.
Deep neural networks (DNNs), in particular, have been used to learn pressure and heat flux closures for fluid models of magnetic reconnection \cite{Laperre2022}, to capture kinetic effects like Landau damping in fluid plasma models \cite{Qin2023, Huang2025, Joglekar2023} as well as in Vlasov-Poisson models using a truncated Fourier-Hermite basis \cite{Barbour2025}, and to efficiently simulate the dynamics of one-dimensional collisional plasmas \cite{Carvalho2024}. Indeed, DNNs have been shown to excel at approximating the complex nonlinear dependencies associated with plasma dynamics, but the inherent complexity of these models hinders their interpretability and often limits their ability to generalize to new conditions outside the training data. These limitations have motivated growing interest in interpretable machine learning techniques, particularly symbolic \cite{BongardLipson2007,SchmidtLipson2009,UdrescuTegmark2020,Cranmer2020} and sparse \cite{Wang2011,Brunton2016,Rudy2017,Schaeffer2017} regression methods. These techniques produce human-readable expressions -- often in the form of partial differential equations (PDEs) - that offer insight into the underlying physics. In plasma physics, sparse regression (SR) has been used to recover the fundamental hierarchy of reduced plasma models [from Vlasov to single-fluid magnetohydrodynamics (MHD)] from fully kinetic plasma simulation data \cite{Alves2022}, to infer fluid moment closures for magnetic reconnection \cite{Donaghy2023} and electrostatic plasma phenomena \cite{Ingelsten2025}, and to infer reduced models of magnetized plasma turbulence \cite{Abramovic2022}; other closely related applications of sparse regression include the development of low-dimensional reduced-order-models for fluid plasma dynamics \cite{Dam2017,Kaptanoglu2021}, recovering the MHD equations from 3D turbulence data \cite{Golden2025}, stellarator design optimization \cite{Kaptanoglu2022,Kaptanoglu2023}, development of plasma profile evolution models near divertors \cite{Lore2023}, and characterization of plasma propulsion dynamics \cite{BayonBujan2024}.

An essential, yet underdeveloped, aspect of ML-based reduced plasma modeling is the incorporation of fundamental physical symmetries. These include, for example, inertial frame invariance (to Lorentz transformations), rotational \& translational invariance, and energy conservation. Embedding these symmetries is crucial to ensure the physical consistency and generalization capabilities of the inferred models \cite{Sanderse2025}. Here we focus on frame invariance. The Vlasov equation, which forms the fundamental mathematical framework for collisionless plasma dynamics, is Lorentz‑invariant \cite{Vlasov1938,Vlasov1967,Hazeltine1998}, and therefore any reduced plasma description derived from Vlasov should respect this symmetry.

% For instance, a physically consistent closure model for the plasma pressure or heat flux tensors should preserve the above symmetries when integrated into the plasma fluid equations. In this work, we highlight Lorentz invariance (which states that the governing equations are invariant to inertial observer) as an important symmetry satisfied by the collisionless Vlasov equation~\cite{Vlasov1938,Vlasov1967,Hazeltine1998}. Embedding these symmetries is crucial to ensure physical consistency and to improve generalization capabilities of the inferred models \cite{Sanderse2025}. 

In general, there are two dominant approaches for embedding symmetries into ML models: (1) hard constraints, which guarantee symmetry preservation, and (2) weak constraints, which encourage, but do not guarantee, symmetry preservation. The first strategy, hard constraints, can be most directly applied in ML by selecting model forms that explicitly satisfy the intended symmetry. For example, by exclusively using symmetry-respecting model terms in SR, the inferred reduced models will preserve the corresponding symmetries, such as rotational and translational invariance~\cite{Gurevich2021}. In the context of plasma physics, low-dimensional POD-Galerkin models for magnetohydrodynamics have been shown to embed global energy conservation by restricting model term nonlinearities to quadratic ~\cite{Kaptanoglu2021}. A more involved method of embedding hard constraints is by building the symmetry into the ML architecture, known as equivariant ML, which has emerged as a powerful paradigm in scientific ML applications \cite{Bronstein2021}. It builds on early work embedding discrete rotational symmetry into convolutional neural networks \cite{Cohen2016}. Recent developments in equivariant ML are beginning to tackle more complex space-time symmetries like Lorentz invariance, for example Lorentz-equivariant architectures for applications in high-energy physics \cite{Spinner2024}, and even Lorentz-equivariant neural PDE solvers for fluid and electrodynamic systems \cite{Zhdanov2024}. While these hard-constrained methods enforce the targeted symmetries, they often come with increased implementation complexity and reduced flexibility in model form.

The second strategy, weak constraints, can be embedded into ML models through data augmentation or loss-based regularization. In loss-based regularization, an additional term in the objective function penalizes deviations from symmetry \cite{Raissi2019,Hao2023}. However, this requires a careful balance between loss coefficients. Data augmentation, on the other hand, incorporates the symmetry by applying symmetry-preserving transformations to the training data, thereby guiding the learned model to respect those transformations \cite{Chen2020}. This method has been widely used in computer vision \cite{Krizhevsky2012,Ciresan2010}, where invariance to translations or rotations is critical. While weak constraints have received less attention in recent ML for physics applications due to the increasing popularity of hard constraint approaches, recent works have demonstrated data augmentation's effectiveness in domains such as particle physics for its ease of embedding complex symmetries. For example, Lorentz-boosted data augmentation has been shown to improve model performance in jet tagging tasks \cite{Monaco2025}.
% Our work revisits this approach in the context of plasma physics, focusing on symmetry-informed data augmentation for the inference of reduced plasma models from first-principles simulations. 

In this work, we revisit data augmentation as a method for embedding physical symmetries in data‑driven plasma models inferred from fully kinetic particle‑in‑cell (PIC) simulations. We specifically focus on the effects of embedding frame-invariance symmetry, according to Lorentz (and Galilean) transformations, in the data-driven inference of reduced plasma models of collisionless magnetic reconnection. In Section \ref{sec:level2:1}, we demonstrate the recovery of the two‑fluid equations for collisionless magnetic reconnection via sparse regression, and show that symmetry‑augmented data (1) yields more accurate coefficients, (2) suppresses spurious symmetry‑violating terms, and (3) reduces the amount of expensive lab‑frame PIC data required. Building on these results, we apply this methodology in Section \ref{sec:level2:2} to the problem of inferring an accurate closure model for the electron fluid pressure tensor in collisionless magnetic reconnection. Specifically, we apply Galilean-boosted data augmentation (the low-velocity limit of Lorentz boosts) to embed Galilean frame-invariance in the closure model, and show that it not only outperforms the model obtained from lab-frame-only data, but also exceeds the accuracy of commonly used analytical closure models. We then further show that symmetry embedding via data augmentation is not limited to SR. Neural networks trained with Galilean-augmented data also yield improved pressure closures, underscoring the broader applicability of this approach. Finally, we conclude in Section~\ref{sec:discuss} by discussing the implications of symmetry embedding for the future development of reduced plasma models capable of faithfully capturing multiscale dynamics.

% In this work, we first recover fluid plasma models of collisionless magnetic reconnection, in the form of the two-fluid equations, from sparse regression with Lorentz-transformed data (Section \ref{sec:level2:1}). We show that incorporating Lorentz-boosted data improves the accuracy of identified model coefficients, eliminates spurious terms that violate Lorentz symmetry, and significantly reduces the amount of lab-frame data required from computationally expensive PIC simulations. Building on this proof of concept, we apply this data-driven discovery framework to the problem of identifying an accurate closure model for the electron fluid pressure tensor in collisionless magnetic reconnection (Section \ref{sec:level2:2}). Specifically, we apply Galilean-boosted data augmentation (the low-velocity limit of Lorentz boosts) to embed Galilean frame-invariance in the closure model. We demonstrate that the pressure closure model obtained from Galilean-augmented data not only outperforms the model obtained from lab-frame-only data, but also exceeds the accuracy of commonly used analytical closure models. Finally, we show that symmetry embedding via data augmentation is not limited to SR: deep neural networks trained with Galilean-augmented data also yield improved pressure closures, underscoring the broader applicability of this approach.

\section{\label{sec:results}Results}

We examine the effectiveness of data-augmentation to embed physical symmetries into machine-learned reduced plasma models from the data of first-principles kinetic simulations. As a case study, we focus on the plasma physics problem of developing reduced fluid plasma descriptions of collisionless magnetic reconnection, which we introduce and motivate below. 
% We then begin by exploring the impact of using data-augmentation to embed frame-invariance (according to Lorentz transformations) of governing fluid plasma equations inferred via sparse regression.

\subsection{\label{sec:level2:1}Recovering Fluid Plasma Models of Magnetic Reconnection from Kinetic Simulation Data}

% We illustrate our symmetry-preserving, sparse regression framework by applying it to the inference of reduced fluid descriptions of collisionless, non-relativistic magnetic reconnection (MR)~\cite{Parker1957,Dungey1953,Sweet1958,Petschek1964,Birn2001} from the data of first-principles, kinetic simulations. 
Magnetic reconnection (MR) \cite{Parker1957,Dungey1953,Sweet1958,Petschek1964,Birn2001} is a fundamental plasma physics process where a change in the magnetic topology converts energy originally stored in the magnetic field to plasma heat, bulk flow kinetic energy, and energetic particle production; it is ubiquitous to both laboratory and astrophysical plasmas. For example, in magnetically confined fusion plasmas magnetic reconnection is responsible for disruptions and sawtooth crashes, while in solar physics MR is responsible for solar flares~\cite{Zweibel2009}. MR is a quintessential multiscale process as there exists a hierarchy of interconnected length and time scales. Changes in the magnetic topology typically occur within a diffusion region at the ion skin depth scale, which is much larger than the electron skin depth and much smaller than the system size~\cite{Liu2022}; a plethora of coherent structures and waves interconnects these disparate scales. Reduced plasma models help solve the multiscale problem by integrating out the finest scale details leading to computational speed up. However, reduced plasma models can also serve a vital role for improving theoretical understanding. Despite significant progress over the last decades, there are still important theoretical challenges in MR, including the origin of the fast reconnection rate ~\cite{Cassak2017,Liu2022} and the dominant mechanisms of anomalous resistivity~\cite{Selvi2023}; the discovery, implementation, and examination of reduced plasma models of MR is important to shed light into these open questions.  

We generate high-fidelity data of the dynamics of collisionless, magnetic reconnection using the fully kinetic PIC simulation code \textit{OSIRIS}~\cite{Fonseca2002,Fonseca2008}. Specifically, we simulate collisionless, low plasma beta (0.1), non-relativistic magnetic reconnection in a pair plasma ($m_e=m_p=1$). To isolate the fundamental plasma process of magnetic reconnection we initialize a doubly-periodic force-free Harris current sheet~\cite{Harris1962} in the $(x,y)$ plane; details about the physical and numerical parameters of the simulation are described in Appendix \ref{app1}. While the spatial grid is in two-dimensions the PIC method evolves particle velocities in all 3 $(x,y,z)$ directions. At each timestep over all spatial grid points we output fluid-moment diagnostics, which are integrals of the particle distribution function with respect to powers of the phase space momentum $\bm{p}$. In general, the $k$th order moment is defined as~\cite{Hazeltine1998}
\begin{equation}\mathcal{M}^{\alpha_1...\alpha_k}\equiv \int \frac{d\bm{p}}{p^0}p^{\alpha_1}...p^{\alpha_k}f
\end{equation}
where $f(\bm{x},\bm{p},t)$ is the particle distribution function. We also output electromagnetic field diagnostics and can construct the electromagnetic field strength tensor $F^{\mu\nu}$. Since the PIC method keeps track of the fluid-moment and electromagnetic fields at different spacetime locations during the loop, we post-process our data by half-time-stepping all particles and spatially centering the electromagnetic field data to ensure all quantities are known at the same spacetime locations. The PIC simulation data used in this paper can be found on Zenodo~\cite{McGraeMenge2025}.
% Following the findings of \textit{Alves \& Fiuza}~\cite{Alves2022}, we post-process our data by integrating the fluid-moments and electromagnetic fields over small space-time volume cubes $\{5\Delta x,5\Delta y,5\Delta t\}$ (which corresponds to physical scales $\{0.5d_e,0.5d_e,0.33w_{pe}^{-1}\}$ respectively) to smooth noise associated with discrete particle fluctuations. The size of these volumes is problem dependent - they must be large enough to smooth the PIC fluctuations yet smaller than characteristic length/times scales in the problem to retain information of physical variation. It was shown that SR equation inference is weakly sensitive to the size of these space-time volumes as long as they meet the two above criteria. 

\subsubsection*{\label{sec:level3:1}Lorentz Data Augmentation}
Our goal is to examine the effects of embedding Lorentz symmetry into reduced fluid models inferred through sparse regression. We accomplish this by augmenting our MR data set using analytic Lorentz transformations (also known as Lorentz boosts). This is the same underlying principle as embedding rotational/translational symmetry by augmenting the data with rotations/translations, with the only difference that the data augmentation operator is Lorentz boosting. For a Lorentz boost velocity $\bm{\beta}=(\beta_x,\beta_y,\beta_z)$ [$c$] (with corresponding Lorentz factor $\gamma_{\bm{\beta}}\equiv\frac{1}{\sqrt{1-\bm{\beta}}^2}$) relative to the simulation, lab-frame we are able to construct the fluid moments as observed by the reference frame moving at velocity $\bm{\beta}$. We denote quantities observed in the Lorentz-boosted reference frame with primed variables,
\begin{equation} \label{Eqn2}
\mathcal{M}^{\alpha'\beta'...\xi'}=\Lambda^{\alpha'}_\mu \Lambda^{\beta'}_\nu...\Lambda^{\xi'}_\rho\mathcal{M}^{\mu \nu ...\rho}
\end{equation}
where $\Lambda^{\mu}_\nu$ is the Lorentz transformation tensor for the given boost $\bm{\beta}$. We construct the electromagnetic fields as seen by the Lorentz moving observer as
\begin{equation}\label{Eqn3}
    F^{\alpha'\beta'}=\Lambda^{\alpha'}_\mu \Lambda^{\beta'}_\nu F^{\mu\nu}
\end{equation}
We also analytically boost temporal and spatial derivatives of the fluid-moments and electromagnetic fields in the lab-frame using analytic Lorentz transformations,
\begin{equation}
    \partial_{\alpha'} =\Lambda^{\mu}_{\alpha'} \partial_{\mu}
\end{equation}
\begin{figure*}[t!]
    \centering
    \includegraphics[width=\textwidth,trim=1
    1 1 1,clip]{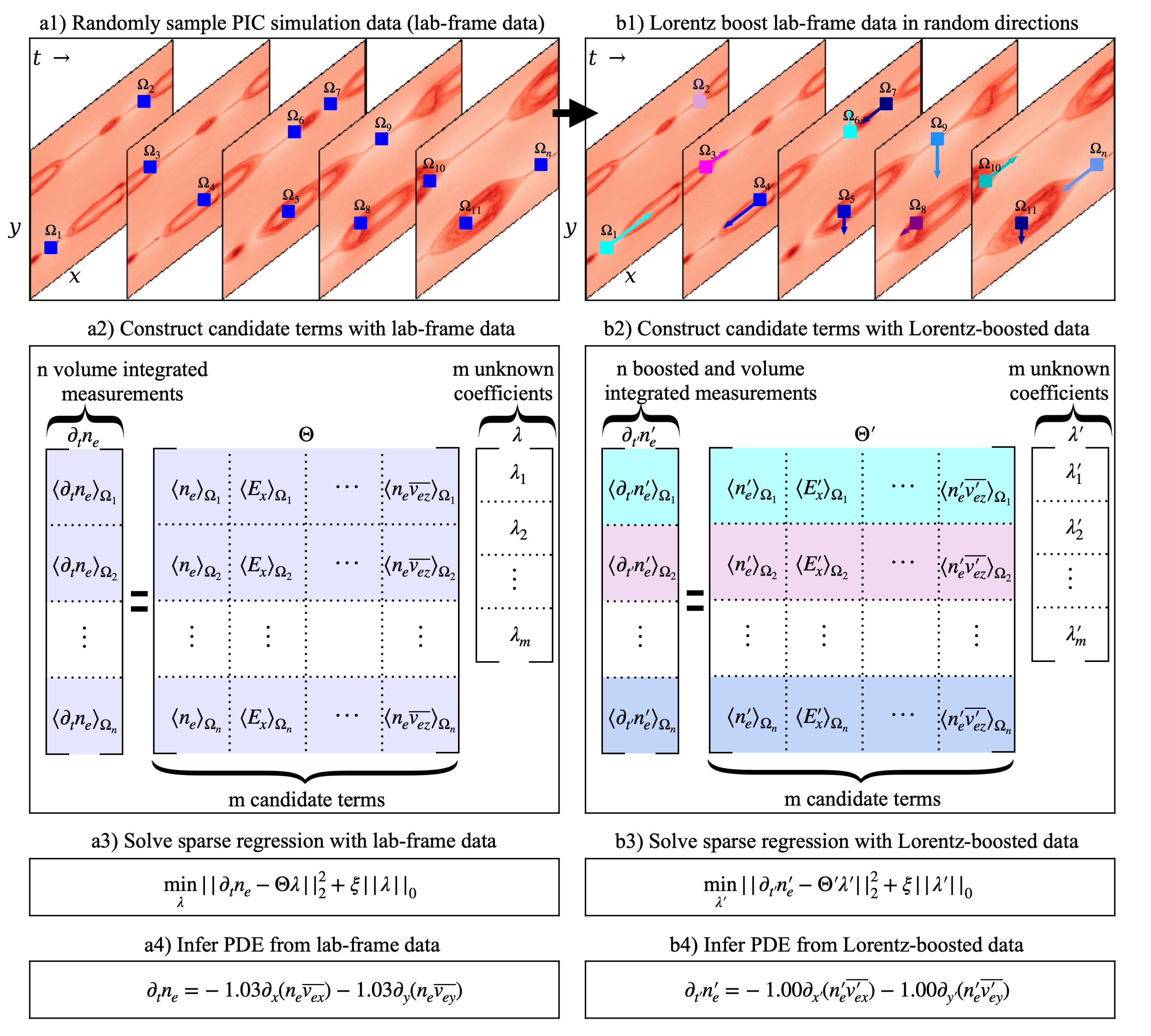}
    \caption{\textbf{Schematic of sparse regression workflow to discover reduced fluid equations with and without Lorentz-transformed data.} A single PIC MR simulation performed in a stationary reference frame (which we denote from now on as the lab-frame) provides lab-frame data for the electron number density $n_e$ (shown above in color), the fluid velocities $\overline{v_{ej}}$, and the electromagnetic fields at finite temporal slices in the $(x,y)$ plane. In step a1) we randomly sample measurements from the lab-frame data, averaging over compact space-time volumes. This averaging smooths noise associated with discrete particle effects. To conduct SR with Lorentz-transformed data, b1), we use analytical transformations to relate the quantities as measured by a random Lorentz observer to the quantities as measured in the simulation, lab-frame. Different color squares represent quantities as measured by Lorentz observers moving with different velocities. In a2) and b2) we balance $m$ measurements of the dynamical term, $\partial_t n_e$ with measurements of catalog terms multiplied by unknown coefficients for lab-frame and boosted data respectively. Crucially, in b2), each Lorentz-boosted reference frame shares the same vector of unknown coefficients $\lambda'$ which embeds the symmetry into the inference procedure. In a3), b3) we solve the SR problem, and in a4), b4) we examine the inferred PDEs which we may compare to the true continuity equation: $\partial_t n_e=-\partial_x(n_e\overline{v_{ex}})-\partial_y(n_e\overline{v_{ey}})$. To infer higher order moment fluid equations, like the momentum or energy equations, we follow the same algorithmic procedure but with different dynamical and catalog terms.}
    \label{fig1}
\end{figure*} 
These Lorentz transformations can all be applied to calculate the physical quantities and their space/time derivatives under any boost from a single lab-frame simulation. In our numerical experiments, we will apply random Lorentz boost transformations of the data along the two principle axes $(x,y)$, forward and backward, for varying Lorentz factors $\gamma_{\bm{\beta}}$.
% in the range $1\leq \gamma_{\bm{\beta}}\leq \gamma_{\bm{\beta }max}=1.5$, which corresponds approximately to $0\leq  |\bm{\beta}| \leq \bm{\beta}_{max}=0.75c$. The maximum fluid velocity in the MR simulation is $\sim 0.5c$, and we choose $\gamma_{\bm{\beta }max}$ large enough such that some Lorentz reference frames experience velocities greater than those in the lab frame. We later explore the effects of choosing different $\gamma_{\bm{\beta }max}$. In the limit $\gamma_{\bm{\beta }max}=1$ there is no data augmentation. All fluid moments and electromagnetic fields at each space-time location are boosted into the Lorentz reference frame moving with corresponding boost velocity $\bm{\beta}$ using Equations \ref{Eqn2} and \ref{Eqn3} [Figure \ref{fig2} (b1)]. We could (and will) combine lab frame data with an arbitrarily large amount of Lorentz boosted data (true data augmentation), however for the sake of a fair comparison we begin by comparing lab data with Lorentz boosted data where each lab measurement is only boosted once.

\subsubsection*{\label{sec:level3:2}Two-Fluid Equations}
To explore the effects of including Lorentz-augmented data on the ML-inference of reduced plasma physics models from kinetic simulation data, we will focus on the recovery of the well-known two-fluid plasma model equations, which we introduce below. The two-fluid equations are Lorentz invariant reduced plasma models and their inference should be improved by embedding Lorentz symmetry. The two-fluid model treats electrons and positrons as separate collisionless fluids. The fluids are coupled together by Maxwell's equations. Each species satisfies an infinite hierarchy of exact, coupled equations which describe the time evolution of the fluid moments $\partial_t\mathcal{M}^{\alpha_1...\alpha_k}$. The equation for the time evolution of the $k\textsuperscript{th}$ moment is derived by taking moments of the collsionless Vlasov equation with respect to $p^{\alpha_1}...p^{\alpha_k}$. Upon truncation, the two-fluid equations provide a substantial computational speed-up compared with PIC as they no longer resolve dynamics of a large number of particles in momentum space as that phase-space coordinate has been integrated over. In this work, we demonstrate our reduced model discovery method on the electron moment equations up to order two: The zeroth order continuity equation (1 equation) which time evolves the time-like component of the four-current density $M^\alpha$,
\begin{equation}\label{Eqn5}
    \partial_t n_e +\partial_{x_j} [n_e \overline{v_{ej}}]=0
\end{equation}
the first order momentum equations (3 equations) which time evolve the time-like components of the stress-energy tensor $\mathcal{M^{\alpha\beta}}$,
% \begin{equation}\label{Eqn6}
%     \partial_t [n_e\overline{p_{ek}}]+\partial_{x_j}[n_e\overline{v_{ej}p_{ek}}]=
% \end{equation}
% \begin{equation}\nonumber
%     -n_e[E_k+\epsilon_{klm}\overline{ v_{el}} B_m]
% \end{equation}
\begin{align}
    \partial_t [n_e\overline{p_{ek}}] &+ \partial_{x_j}[n_e\overline{v_{ej}p_{ek}}]= \nonumber
    \\ &-n_e[E_k+\epsilon_{klm}\overline{ v_{el}} B_m]
\end{align}
and the second order energy equations (6 unique equations) which time evolve the time-like components of the stress-flow tensor $M^{\alpha\beta\gamma}$,
% \begin{equation}\label{Eqn7}
%     \partial_t [n_e\overline{ p_{ek}p_{el}}]+\partial_{x_j}[n_e\overline{ v_{ej}p_{ek}p_{el}}]=
% \end{equation}
% \begin{equation}\nonumber
%     -n_e[\overline{ p_{ek}} E_l+\overline{ p_{el}} E_k
% \end{equation}
% \begin{equation}\nonumber
%     +\epsilon_{lmn}\overline{ p_{ek}v_{em}} B_{n}+\epsilon_{kop}\overline{ p_{el}v_{eo}} B_{p}]
% \end{equation}
\begin{align}
    \partial_t [n_e\overline{ p_{ek}p_{el}}] & +\partial_{x_j}[n_e\overline{ v_{ej}p_{ek}p_{el}}]= \nonumber
    \\ & -n_e[\overline{ p_{ek}} E_l+\overline{ p_{el}} E_k \nonumber
    \\ & +\epsilon_{lmn}\overline{ p_{ek}v_{em}} B_{n}+\epsilon_{kop}\overline{ p_{el}v_{eo}} B_{p}]
    \label{Eqn7}
\end{align}
 where $n_e$ is the electron density and $\overline{a}\equiv \frac{1}{n_e} \int f_e a dp_e^3$. $|q_e|=m_e=1$ in normalized units. $v_{ej}$ is the electron phase space velocity in the $j$th direction which is related to the electron phase space momentum $p_{ej}=\frac{v_{ej}}{\sqrt{1-v_{ex}^2-v_{ey}^2-v_{ez}^2}}$. The 1 continuity equation, 3 momentum equations, and 6 energy equations comprise 10 moment equations which we attempt to rediscover. In principle, we could recover moment equations higher than order two, however we truncate our demonstration at the level of the energy equations. 
 
\subsubsection*{\label{sec:level3:3}Embedding Lorentz Invariance into the Sparse Regression Recovery of Two-Fluid Equations}

To recover the two-fluid equations from our collisionless magnetic reconnection data we follow the algorithmic procedure as proposed in Ref.~\cite{Rudy2017} PDE-FIND (the PDE generalization of SINDy~\cite{Brunton2016}). PDE-FIND discovers partial differential equations directly from data by using sparse regression to fit unknown coefficients next to a prescribed basis of catalog terms to describe a target term. SR differs from traditional regression as it penalizes non-zero coefficients, resulting in a sparse approximation to the target. This leads to PDEs which capture the essence of the dynamics while still being interpretable. In this study the regression targets are the time derivative terms from the two-fluid continuity, momentum, and energy equations as expressed in equations \ref{Eqn5}-\ref{Eqn7}. Hence, we have ten unique regression targets and perform the SR equation inference procedure ten separate instances. For each equation inference we provide a basis comprised of fluid-moments, electromagnetic fields, their spatial derivatives, and polynomial combinations of the aforementioned terms in an attempt to describe the target dynamical term; see Appendix \ref{app3} for a full description of the catalog terms for each equation inference.

Importantly, to effectively handle the evaluation of PDE terms in the presence of noisy data, associated with discrete particle fluctuations (inherent to PIC simulation data), we pose the SR inference of PDEs in their integral form rather than their differential form \cite{Crutchfield1987,Schaeffer2017_2,Reinbold2020,Messenger2021,Alves2022}. This technique was shown to be crucial for the accurate and robust inference of plasma PDEs from the data of first-principles PIC simulations \cite{Alves2022}, and has become well established in recent years ~\cite{Reinbold2020,Messenger2021}. Thus, in this work the target time derivative terms and the library of candidate PDE terms are evaluated using finite differences on the data, followed by integrating over small compact space-time volumes $\{5\Delta x,5\Delta y,5\Delta t\}$ (which corresponds to physical scales $\{0.5d_e,0.5d_e,0.33w_{pe}^{-1}\}$). In general, a suitable choice for the size of these volumes is problem dependent - they should be large enough to smooth the PIC fluctuations, yet smaller than characteristic length/times scales in the problem to retain information of physical variation. In our numerical experiments here, the choice of $\{5\Delta x,5\Delta y,5\Delta t\}$ volumes satisfies the above criteria for collisionless magnetic reconnection. We further confirmed that our results were weakly sensitive to the size of these space-time volumes, as long as they met the two above criteria. 
\begin{figure*}[t!]
    \centering
    \includegraphics[width=\textwidth]{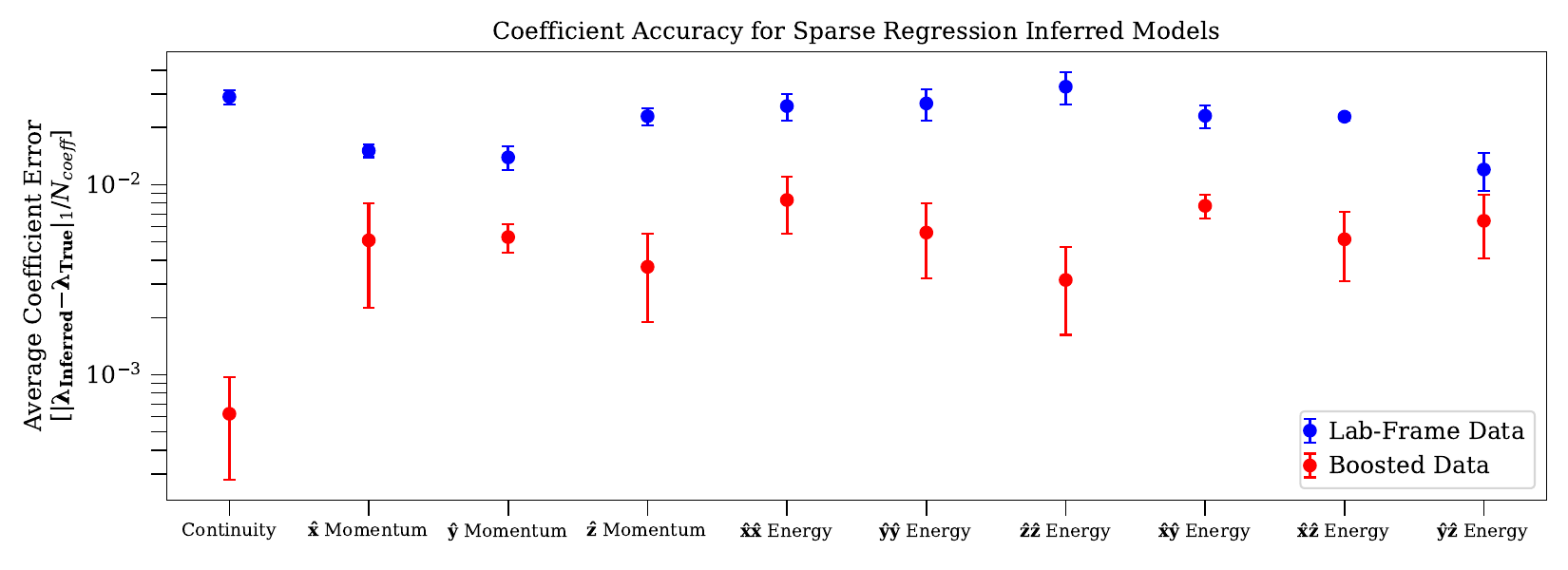}
    \caption{\textbf{Including Lorentz-boosted data in the re-discovery of the two-fluid reduced model equations greatly improves coefficient accuracy.} On the vertical axis is average model coefficient error measured as the L1 norm between true and inferred coefficients normalized by the number of non-time derivative terms in the equation, while on the horizontal axis are the first ten unique two-fluid plasma equations. The blue points represent coefficient errors for SR inferred equations with only lab-frame data, while red points represent boosted data. The vertical error bars correspond to $\pm$ one standard deviation in average coefficient error, and the statistics come from training each model five times with different realizations of the random data sampling. For all ten equations Lorentz boosting improves the mean accuracy of the inferred coefficients beyond one standard deviation.}
    \label{fig2}
\end{figure*}
For each equation inference we begin by performing SR on only the original lab-frame data, without Lorentz transformations. We use a sample of $15^3=3375$ randomly distributed $\{5\Delta x,5\Delta y,5\Delta t\}$ measurement volumes from the full MR data set [Figure \ref{fig1} (a1)].  In each of these sampled volumes we evaluate the target terms (for example $\partial_t n_e$ which appears in the continuity equation, and the library of catalog terms [Figure \ref{fig1} (a2)]) and integrate them over each volume; see Appendix \ref{app2} for more data sampling details. Each of the catalog terms is attached with an unknown coefficient $\bm{\lambda}$. All measurements of a catalog term share the same unknown coefficient. SR then fits $\bm{\lambda}$ to the data in an \textit{optimal} manner [Figure \ref{fig1} (a3)] using a variation of the sequential thresholding least-squares algorithm as proposed in Ref.~\cite{Brunton2016} and ten-fold cross validation, and we infer a governing equation from the regression procedure [Figure \ref{fig1} (a4)]. A detailed description of the SR algorithm used in this work is described in Ref.~\cite{Alves2022}.

% For each equation inference we begin by performing SR on only the original lab frame data without Lorentz transformations. From our 3,375 sampled measurements we construct the target terms (for example $\partial_t n_e$ which appears in the continuity equation) as well as the library of catalog terms [Figure \ref{fig2} (a2)]. Each of the catalog terms is attached with an unknown coefficient $\bm{\lambda}$. All measurements of a catalog term share the same unknown coefficient. SR then fits $\bm{\lambda}$ to the data in an \textit{optimal} manner [Figure \ref{fig2} (a3)] using a variation of the sequential thresholding least-squares algorithm as proposed in Ref.~\cite{Brunton2016} and tenfold cross validation, and we infer a governing equation from the regression procedure [Figure \ref{fig2} (a4)]. For more details on the SR algorithm and how it determines coefficients in a sparsity promoting manner see Ref.~\cite{Alves2022}. 

We now repeat the PDE discovery procedure using Lorentz-transformed data. Using the analytical Lorentz transformations described earlier, we boost each sampled volume into a randomly moving Lorentz reference frame as illustrated in Figure~\ref{fig1} (b1). Each volume is boosted in a random direction (along the two principal $(x,y)$ axes), with a random Lorentz factor sampled uniformly in the range $1\leq \gamma_{\bm{\beta}}\leq \gamma_{\bm{\beta }max}=1.5$, which corresponds approximately to $0\leq  |\bm{\beta}| \leq \bm{\beta}_{max}=0.75c$. The maximum fluid velocity in the MR simulation is $\sim 0.5c$, and we choose $\gamma_{\bm{\beta }max}$ large enough such that some Lorentz reference frames experience velocities greater than those in the lab-frame; we will later examine the effects of choosing different $\gamma_{\bm{\beta }max}$. Each row of the candidate term matrix now corresponds with an observation of the measured target and catalog terms taken by a randomly moving Lorentz observer, as shown in Figure \ref{fig1} (b2). While we later consider combining lab-frame data with arbitrarily many Lorentz-boosted samples for true data augmentation, we begin by comparing lab-frame-only and boosted-only datasets, with each lab-frame sample boosted once.

Having carried out the random Lorentz transformations of each measurement volume, we construct our framework to ensure that all Lorentz reference frames agree on the same coefficients $\bm{\lambda'}$, which embeds Lorentz invariance into SR. With the Lorentz-transformed data we then solve the regression procedure [Figure \ref{fig1} (b3)] and infer the governing PDE [Figure \ref{fig1} (b4)]. We repeat this procedure for each of the ten two-fluid equations, extracting model coefficients discovered with lab-frame-only and boosted-only data. In the following, we discuss the effects of using Lorentz-transformed data on the inferred models.
\begin{figure}[t!]
    \centering
    \includegraphics[width=0.5\textwidth]{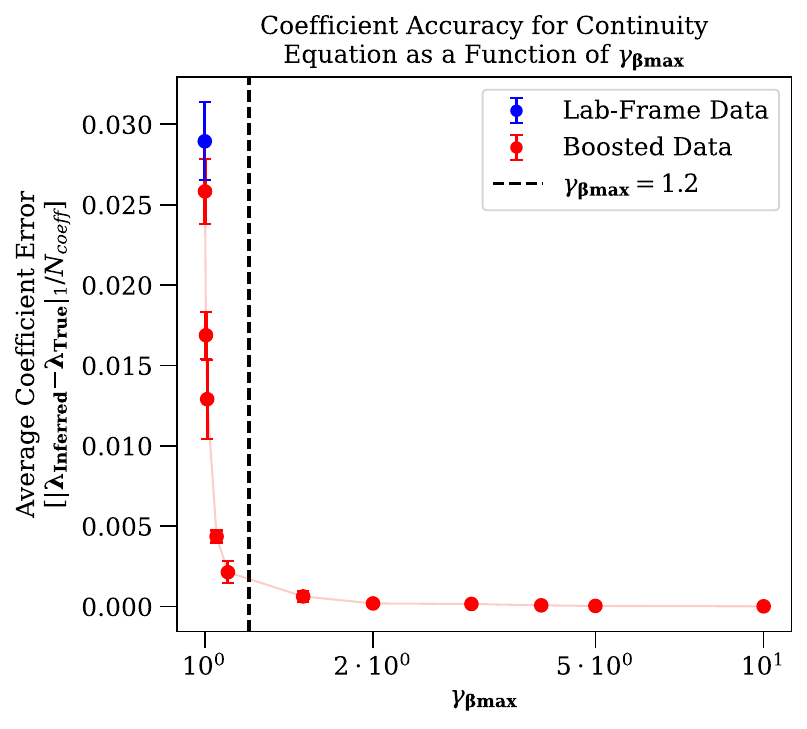}
    \caption{\textbf{Average coefficient error decreases as a function of maximum Lorentz boost ($\gamma_{\bm{\beta}max}$).} The vertical axis measures the L1 difference between true and SR inferred coefficients. We randomly Lorentz transform our data in the range $1\leq \gamma_{\bm{\beta}}\leq\gamma_{\bm{\beta}max}$, and demonstrate that increasing $\gamma_{\bm{\beta}max}$ (embedding the symmetry more stringently) leads to more accurate coefficients for the continuity equation recovery. The vertical error bars correspond to $\pm$ one standard deviation in average coefficient error, and the statistics come from training each model five times with different realizations of the random data sampling.}
    \label{fig3}
\end{figure}

\begin{figure*}[t!]
    \centering
    \includegraphics[width=0.95\textwidth]{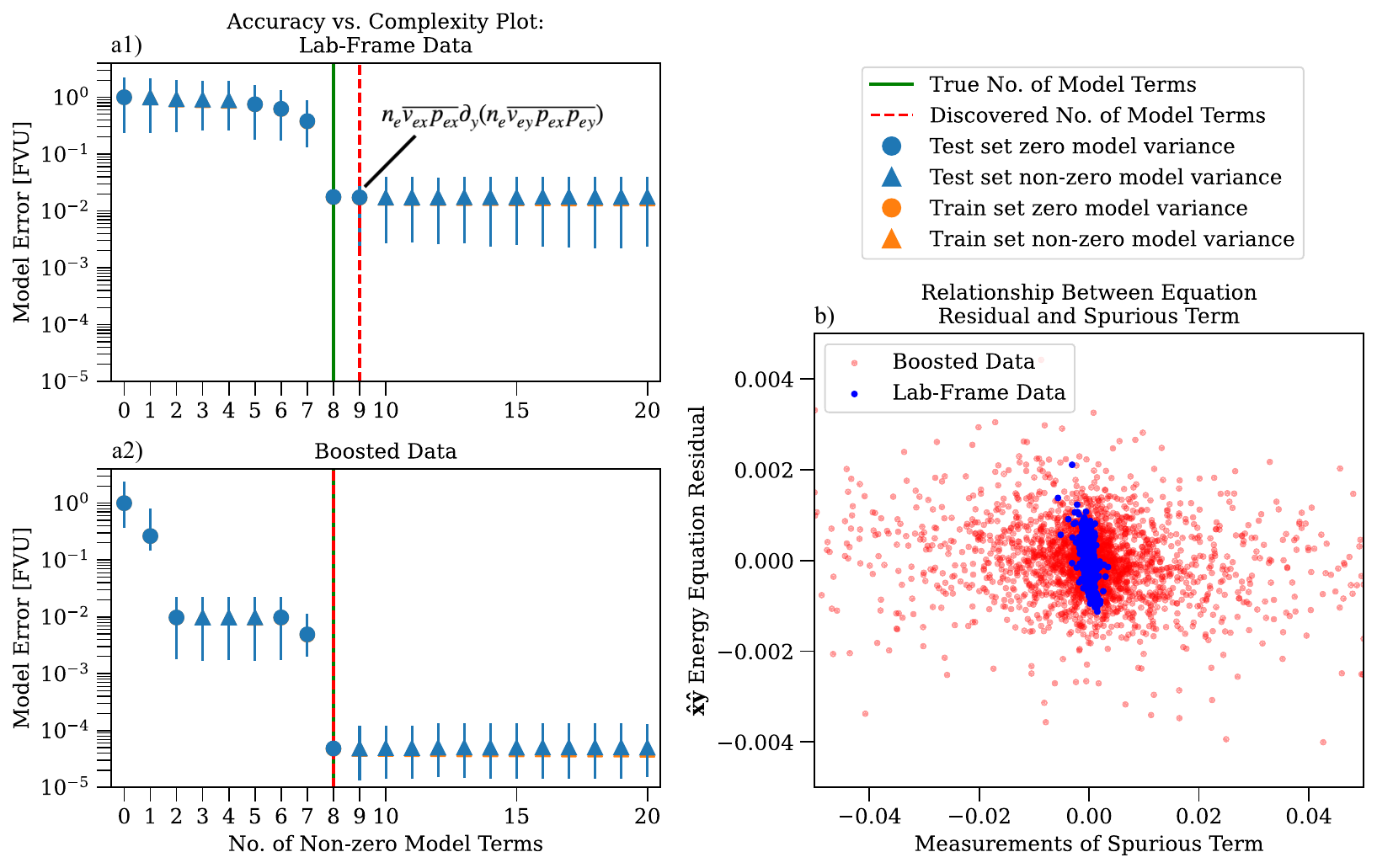}
    \caption{\textbf{Lorentz data transformation removes an unwanted, spurious model term identified by the lab-frame sparse regression.} In plots a1), lab-frame data, and a2), Lorentz-boosted data, we display the sparse regression inferred $\bm{\hat{x}\bm{\hat{y}}}$ energy equation model error as a function of model terms. Circles represent zero model form variance for all SR K-Folds, while triangles represent non-zero. See the upper right quadrant for the legend for plots a1) and a2). Accuracy is measured in terms of Fraction of Variance Unexplained (FVU). FVU$\equiv \frac{Var(\hat{f}-f)}{Var(f)}$ where $f$ is the target variable (in this case measurements of $\partial_t[n_e\overline{ p_{ex}p_{ey}}]$) and $\hat{f}$ is the model (in this case the SR inferred model for $\partial_t[n_e\overline{ p_{ex}p_{ey}}]$). Complexity is measured in terms of the number of non-zero model terms in the equation. The true $\bm{\hat{x}\bm{\hat{y}}}$ energy equation has 8 terms and is marked with a green vertical line. SR performed with lab-frame data identifies an incorrect 9 term model (an extra incorrect term), while with boosted data identifies the correct 8 term model; these models are marked by red vertical lines. Plot b) offers an explanation for Lorentz boosting removing the extra, spurious term; a once linear relationship between measurements of the spurious 9th term ``$n_e\overline{ v_{ex}p_{ex}} \partial_y[n_e\overline{ v_{ey}p_{ex}p_{ey}}]$'' and the $\bm{\hat{x}\bm{\hat{y}}}$ energy equation residual breaks upon Lorentz boosting. The relation becomes more non-linear and disperse. With lab-frame data the Pearson Correlation Coefficient (PCC), which measures normalized covariance of two variables, is -0.399. For boosted data the PCC is -0.204 meaning that the linear relationship lessened upon boosting. }
    \label{fig4}
\end{figure*}

\subsubsection*{\label{sec:level3:4}Improved Coefficient Accuracy}

The first and most apparent finding is that the coefficients inferred from SR are more accurate when using Lorentz-transformed data. Since the two-fluid equations can be derived analytically by taking moments of the collisionless Vlasov equation, we can compare the true two-fluid coefficients with the inferred coefficients [Figure \ref{fig2}]. For all ten two-fluid reduced model equations, incorporating randomly Lorentz-boosted data in the training set decreases average coefficient error beyond a standard deviation. The models discovered with lab-frame data have an average coefficient error of 1.34\%, while for the models discovered with boosted data it is 0.15\%. Relatively, the boosted models are 9 times more accurate than lab-frame models. See Appendix \ref{app5} for a table of discovered coefficients used to create Figure \ref{fig2}. 

We also examine how coefficient accuracy improves as a function of $\gamma_{\bm{\beta}max}$, the maximum Lorentz factor of each random Lorentz boost. Figure \ref{fig3} shows coefficient accuracy as a function of $\gamma_{\bm{\beta}max}$ for the inferred continuity equation; we observe an elbow in the accuracy vs. boost magnitude plot at $\gamma_{\bm{\beta}max}\sim 1.2$, corresponding to a boost magnitude $|\bm{\beta}|\sim 0.5c$, which is the maximum fluid proper velocity reached in the simulation. To reap the full benefits of Lorentz data transformations one must Lorentz boost data into regimes outside of the existing dynamics. We show additional examples of average coefficient error decreasing with increasing $\gamma_{\bm{\beta}max}$ for the $\bm{\hat{z}}$ momentum and $\bm{\hat{z}}\bm{\hat{z}}$ energy equations in Appendix \ref{app4}.

To shed light into why coefficient accuracy improves as a function of $\gamma_{\bm{\beta}max}$, let us write down the regression loss function for the inference of the continuity equation with data that has been Lorentz transformed by $\bm{\beta}=\beta \bm{\hat{x}}$,
\begin{equation}\label{Eqn8}
\mathcal{L'}\sim\partial_{t'}n_{e}'+\lambda_1'\partial_{x'}[n_e'\overline{ v_{ex}'}]
\end{equation}
where $\lambda_1'$ is an unknown coefficient. From the Lorentz-boosted simulation data we have measurements of the terms $\partial_{t'}n_{e}'$ and $\partial_{x'}[n_e'\overline{ v_{ex}'}]$. Using the inverse Lorentz transform $(\Lambda^{-1})^{\mu}_\nu$ we can express $\mathcal{L}'$ in terms of the lab-frame moments. Crucially we define the unknown coefficient $\lambda_1'=\lambda_1$ to be invariant under the inverse Lorentz transformation.
% \begin{equation}\label{Eqn9}
%     \mathcal{L'}\sim \partial_t n_e+\lambda_1\partial_{x}[n_e\overline{ v_{ex}}]+
% \end{equation}
% \begin{equation}\nonumber
%     \beta \{[\lambda_1-1][\partial_{t}(n_e\overline{ v_{ex}})-\partial_xn_e]-\beta[\lambda_1\partial_t n_e+\partial_{x}(n_e\overline{ v_{ex}})]\}
% \end{equation}
\begin{align}
    \mathcal{L'}\sim \partial_t n_e &+\lambda_1\partial_{x}[n_e\overline{ v_{ex}}] 
    \nonumber
    \\ &+ \beta \{[\lambda_1-1][\partial_{t}(n_e\overline{ v_{ex}})-\partial_xn_e]
    \nonumber
    \\ & -\beta[\lambda_1\partial_t n_e+\partial_{x}(n_e\overline{ v_{ex}})]\}
    \label{Eqn9}
\end{align}
We recognize the first line of Equation \ref{Eqn9} to be the lab-frame loss function $\mathcal{L}\sim \partial_t n_e+\lambda_1\partial_{x}[n_e\overline{ v_{ex}}]$.   
% \begin{equation}\label{Eqn10}
%     \mathcal{L'}\sim \mathcal{L}+
% \end{equation}
% \begin{equation}\nonumber
%     \beta \{[\lambda_1-1][\partial_{t}(n_e\overline{ v_{ex}})-\partial_xn_e]-\beta[\lambda_1\partial_t n_e+\partial_{x}(n_e\overline{ v_{ex}})]\}
% \end{equation} 
\begin{align}\label{Eqn10}
    \mathcal{L'}\sim \mathcal{L} & +
    \beta \{[\lambda_1-1][\partial_{t}(n_e\overline{ v_{ex}})-\partial_xn_e] \nonumber 
    \\ &-\beta [\lambda_1\partial_t n_e+\partial_{x}(n_e\overline{ v_{ex}})]\}
\end{align} 
Equation \ref{Eqn10} tells us the boosted frame loss function is proportional to the lab-frame loss function plus additional terms tuned by the boost magnitude $\beta$. In the limit $\gamma_{\bm{\beta}max}\rightarrow 1$, corresponding to $\beta\rightarrow0$, $\mathcal{L'}\sim\mathcal{L}$. Much more interesting is the large Lorentz boost limit $\gamma_{\bm{\beta}max}\rightarrow \infty$, corresponding to $\beta\rightarrow 1$. In this limit $\beta^2= 1+2(\beta-1)+\mathcal{O}[(1-\beta)^2]$ and Equation \ref{Eqn9} becomes
\begin{equation}\label{Eqn11}
    \mathcal{L'}\sim [1-\lambda_1]\{\partial_t[n_e-n_e\overline{ v_{ex}}]+\partial_x[n_e-n_e\overline{ v_{ex}}]\}
\end{equation}
Measurements of the term $\{\partial_t[n_e-n_e\overline{ v_{ex}}]+\partial_x[n_e-n_e\overline{ v_{ex}}]\}$ from Equation \ref{Eqn11} are not typically equal to zero in the MR data set, thus to minimize the loss function in the large Lorentz boost limit the multiplicative term $[1-\lambda_1]\rightarrow0$, $\lambda_1\rightarrow 1$. From theory we know that the true continuity coefficient is indeed $\lambda_1=1$. In the same manner boosting along the $\bm{\hat{y}}$ axis would fix the unknown coefficient attached to $\partial_{y'}[n_e'\overline{ v_{ey}'}]$ to its true value, also $1$. This line of analysis tells us that Lorentz transforming our dataset combined with fixing the unknown coefficients to be the same in all Lorentz reference frames non-trivially enforces the true coefficient value into the equation discovery. We further show in Appendix \ref{app6} that not only are these coefficients more accurate, but they lead to models which better satisfy Lorentz invariance.

\subsubsection*{\label{sec:level3:6}Elimination of Spurious Unphysical Terms}
ML models find correlations in the data, but these correlations may not be physical. We find that Lorentz data augmentation can break unphysical correlations in ML models. For example, when discovering the $\bm{\hat{x}}\bm{\hat{y}}$ component of the energy equation with only lab-frame MR data, an unphysical term is consistently identified as the 9th model term by the lab-frame SR, even though we know from theory that the true equation has 8 terms [Figure \ref{fig4} (a1)]. However, upon boosting our MR data into randomly moving Lorentz reference frames, we see in Figure \ref{fig4} (a2) that the regression identifies the correct 8 term model without the spurious term. Figure \ref{fig4} (b) reaveals how the spurious term is removed by boosting: in the lab-frame, there exists a linear correlation between a spurious candidate term $n_e\overline{ v_{ex}p_{ex}} \partial_y[n_e\overline{ v_{ey}p_{ex}p_{ey}}]$ and the model residual. Hence, it is convenient for the regression procedure to attach a nonzero coefficient, $-0.2298$, to this term which reduces the loss function. Yet, after applying Lorentz boost transformations to the data, the once linear relation between model loss and term measurement is broken. This is because the functional form of the spurious term is not Lorentz invariant and appears different for independent Lorentz moving observers, i.e.
\begin{figure}[t!]
    \includegraphics[width=0.5\textwidth]{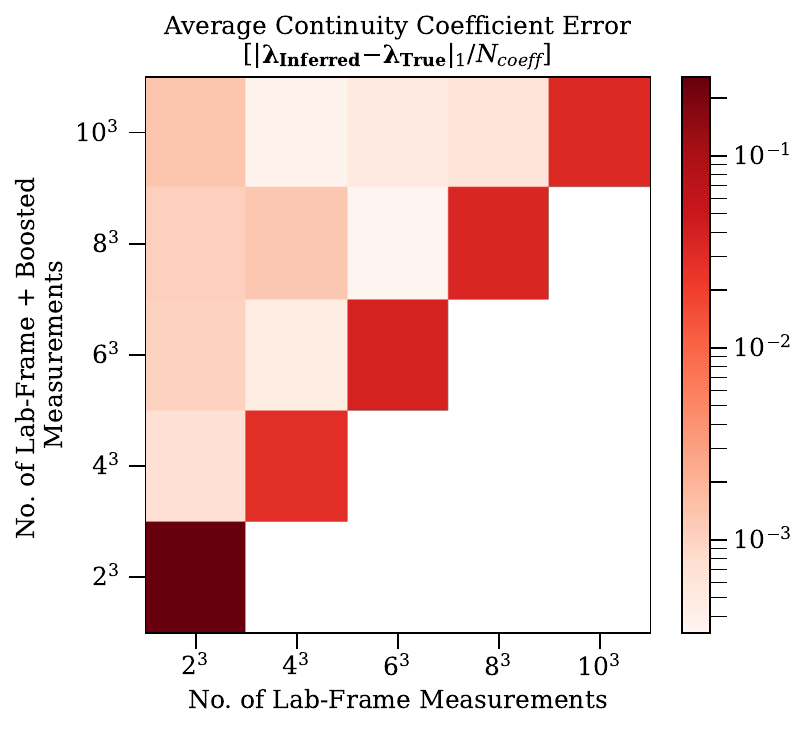}
    \caption{\textbf{Lorentz boosting improves data efficiency in reduced model discovery.} The color is the average continuity coefficient error measured with the L1 norm. The horizontal axis depicts the number of randomly sampled lab-frame measurements from our MR simulation. The vertical axis represents this number of lab-frame measurements \textit{plus} random Lorentz augmentations of those lab-frame measurements. The identity line corresponds to no data augmentation. For a fixed horizontal position corresponding to a fixed number of lab-frame data samples, moving vertically corresponds to increasing the number of Lorentz augmentations of those original, lab-frame samples. We see a sharp drop in average continuity coefficient error moving off of the identity line, indicating that embedding Lorentz invariance into the SR equation inference has more of an effect on coefficient accuracy than simply including more lab-frame data. Note that the continuity equation has two model terms in two dimensions thus the normalizing factor $N_{coeff}=2$ in this case.}
    \label{fig5}
\end{figure} 
$n_e\overline{ v_{ex}p_{ex}} \partial_y[n_e\overline{ v_{ey}p_{ex}p_{ey}}]\neq n_e'\overline{ v_{ex}'p_{ex}'} \partial_{y'}[n_e'\overline{ v_{ey}'p_{ex}'p_{ey}'}]$. This is perhaps the most important effect of Lorentz data augmentation, and is crucial to enable generalizable ML models. We will return to this point in Section \ref{sec:level2:2}.

\subsubsection*{\label{sec:level3:7}Improved Data Efficiency}

State-of-the-art machine learning models are notoriously data-hungry~\cite{Adadi2021}. Generating, storing, pre-processing, transferring, and training with large amounts of data can be a significant bottleneck in the development of generalizable ML reduced models of physical systems. Here, we find that not only does Lorentz data augmentation increase the amount of physically consistent training data without having to generate more through expensive PIC simulations, but we find it improves the model accuracy significantly faster than by including more original, lab-frame data. We show this effect in Figure \ref{fig5}, where we examine coefficient accuracy for the continuity equation as a function of varying the relative proportions of original lab-frame data and Lorentz-transformed data. With only $2^3=8$ lab-frame measurements, in the lower left corner of Figure \ref{fig5}, we find a large average continuity coefficient error of $0.257$. This is not surprising given the small amount of data used to build the model. Moving up one square, we randomly boost these $8$ lab-frame measurements $56$ times to have $4^3=64$ total lab-frame + boosted measurements and the average continuity coefficient error significantly decreases to $0.001$. This same phenomenon is observed for any starting number of lab-frame measurements. For a fixed number of lab-frame samples, the coefficient error saturates after a sufficient number of augmented boosted sample. Interestingly, even with $10^3=1,000$ lab-frame measurements we observe an average continuity coefficient error of $0.032$, much higher than what can be achieved with only $2^3=8$ lab-frame measurements combined with Lorentz boosts. These results underscore the effectiveness of data augmentation in improving the accuracy of ML models.

\subsection{\label{sec:level2:2}Discovering Symmetry-Preserving Fluid Closures}
\subsubsection*{\label{sec:level3:8}Sparse Regression Inferred Closure}
\begin{figure*}[t!]
    \centering
    \includegraphics[width=\textwidth]{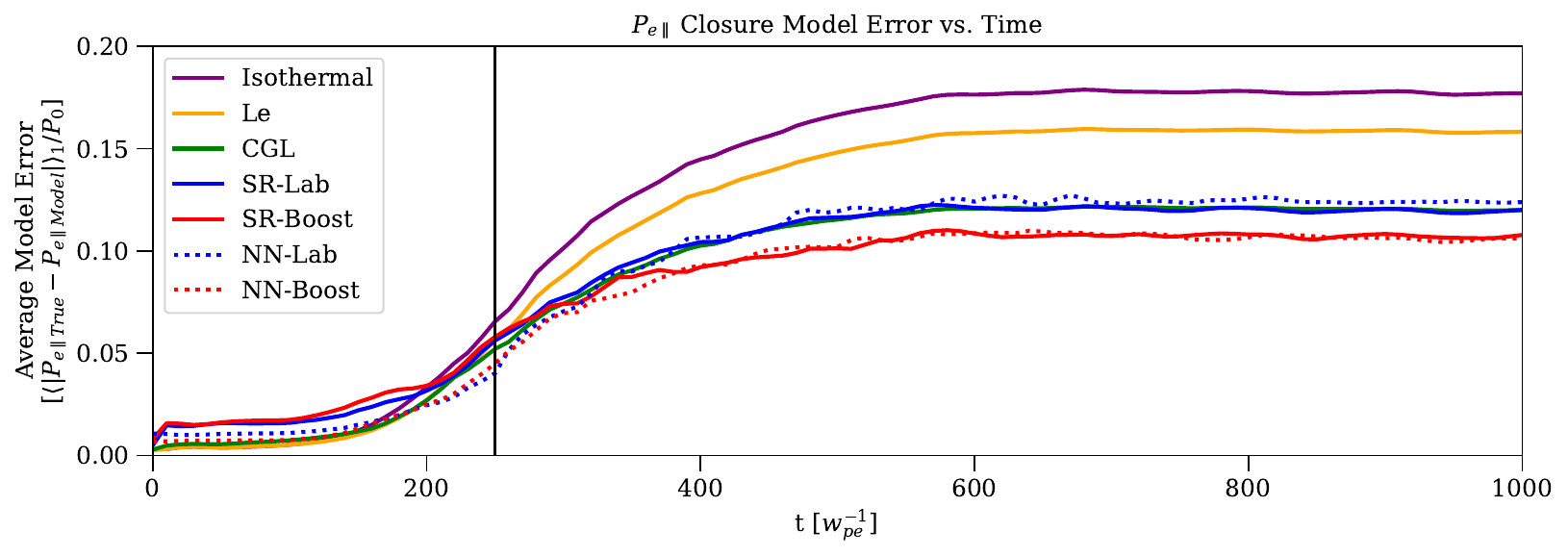}
    \caption{\textbf{Machine learned $P_{e\parallel}$ pressure closures for collisionless magnetic reconnection - discovered with Galilean-augmented data - not only outperform models discovered without symmetry embedding, but further commonly used analytic pressure closures.} The vertical axis corresponds to model error measured as the spatial mean of the absolute L1 difference between the true simulation pressure and the pressure closure model. We normalize the average pressure error to a characteristic pressure value in the simulation $P_0=0.1 n_{0}m_ec^2$. The horizontal axis is time. The sparse regression (SR) and neural network (NN) models are trained until $t=t_{max}=250w_{pe}^{-1}$ and then generalized to future times. Isothermal, Le, and CGL are commonly used pressure models with analytic functional forms. As time increases and reconnection becomes more non-linear and complicated all closures perform worse. At late, generalization times the machine learned closures discovered with Galilean-augmented data have the lowest model error.}
    \label{fig6}
\end{figure*} 
Having established the benefits of Lorentz-boosted data augmentation on the recovery of known equations in Section \ref{sec:level2:1}, we now extend the application of this technique to the development of accurate, interpretable, and symmetry-preserving plasma fluid closures from our kinetic simulations of collisionless magnetic reconnection. The development of such closure models is important to enable accurate and computationally efficient fluid modeling of magnetic reconnection in multi-scale plasma scenarios, such as magnetized plasma turbulence and magnetospheric physics \cite{Le2009,Ohia2012,Ng2017,Ng2020}.
% In addition to enabling computationally efficient modeling of multi-scale plasma dynamics, the development of such closure models may also shed new physical insights into the interplay between kinetic effects and plasma dynamics at large fluid scales. 

Here, we focus on the data-driven inference of a closure model for the electron pressure tensor for collisionless magnetic reconnection. Specifically, we seek to express the electron pressure tensor ${P_{ejk}}\equiv\int f_e ({v_{ej}}-\overline{{v_{ej}}})({p_{ek}}-\overline{{p_{ek}}}) dp_e^3$ as a function of lower order moments and fields to close the infinite hierarchy of two-fluid equations introduced in Section \ref{sec:level2:1}. While the pressure tensor appears in a Lorentz-invariant equation, it itself is not Lorentz-invariant. However, for small Lorentz-boosts (Galilean boosts) the pressure tensor remains invariant and thus we should expect our SR inferred closures to abide by this important symmetry. In the large magnetic field limit (applicable in guide field reconnection) the fluid moments and electromagnetic fields transform as follows under Galilean transformations, where $\beta_j$ is a Galilean boost velocity vector and quantities as observed in the moving reference frame are denoted by primed variables:
% \begin{equation}
%     n_e' = n_e
% \end{equation}
% \begin{equation}
%     \overline{ {v_{ej}'}} = \overline{ {v_{ej}}} - \beta_j
% \end{equation}
% \begin{equation}\label{eqn18}
%     {P_{ejk}'}={{P_{ejk}}}
% \end{equation}
% \begin{equation}
%     E_j'=E_j+\epsilon_{jkl}\beta_k B_l
% \end{equation}
% \begin{equation}\label{eqn20}
%     B_j'=B_j
% \end{equation}
\begin{align}
    n_e' &= n_e
    \\ \overline{ {v_{ej}'}} &= \overline{ {v_{ej}}} - \beta_j
    \\ {P_{ejk}'}&={{P_{ejk}}} \label{eqn18}
    \\ E_j'&=E_j+\epsilon_{jkl}\beta_k B_l
    \\B_j'&=B_j  \label{eqn20}
\end{align}

Instead of discovering the pressure tensor component by component in a Cartesian basis, the more natural basis in magnetized plasmas is defined with respect to the local magnetic field. In this basis, the relevant components are the parallel ($P_{e\parallel}$) and perpendicular ($P_{e\perp}$) components of the pressure tensor. %Indeed, we have verified in the PIC simulation data of the MR regime considered in this work that the pressure tensor is nearly gyrotropic, i.e. $\mathbf{P_e} \simeq P_{e\parallel} \mathbf{b}\mathbf{b} + P_{e\perp} (\mathbf{I}-\mathbf{b}\mathbf{b})$, where $\mathbf{I}$ is the identity tensor, and $\mathbf{b}\equiv\mathbf{B}/|\mathbf{B}|$.

To illustrate how data augmentation improves closure discovery, we focus on the component of the pressure tensor parallel to the local magnetic field. From Equations \ref{eqn18} and \ref{eqn20} we see that the electron pressure local to the magnetic field is invariant under a Galilean boost, thus for any Galilean boost $\beta_j$, $P_{e\parallel}'=P_{e\parallel}$ and $P_{e\perp}'=P_{e\perp}$.
% \begin{equation}
    % P_{e\parallel}'=P_{e\parallel}
% \end{equation}
% In the same manner, $P_{e\perp}'=P_{e\perp}$. 
The evolution of $P_{e\parallel}$ is more complex than the evolution of $P_{e\perp}$ in the MR regime considered in this work (medium guide field regime), thus we focus here in the main text on discovering a closure model for $P_{e\parallel}$; we discuss closure model discovery for $P_{e\perp}$ in Appendix \ref{app7}. 

We sample local and instantaneous quantities at $10^3$ spatial locations every $10w_{pe}^{-1}$ from $t_{min}=0w_{pe}^{-1}$ to $t_{max}=250w_{pe}^{-1}$; $t_{max}$ also corresponds to the time at which $\sim 50\%$ of the initial magnetic flux has been reconnected. In the same spirit as the two-fluid equation inference we only sample from spatial locations where the target variable, in this case $P_{e\parallel}$, is greater than its $70$th percentile value at its sampled timestep. We observe how well our pressure closure generalizes to times outside of the training window, $t>t_{max}$. We provide a library of candidate terms comprised of fluid-moments and electromagnetic fields $\{n_e,\overline{{v_{ex}}},\overline{{v_{ey}}},\overline{{v_{ez}}},B_x,B_y,B_z\}$ and their polynomial combinations; the electric field is left out as a candidate term due to its strong correlation with $\overline{\bm{v_e}}\times\bm{B}$.

Without Galilean-boost augmented data, SR infers the following seven-term electron parallel pressure model (which we refer to as the SR-Lab model): 
% \begin{equation}\label{eqn22}
% P_{e\parallel lab}=
% -0.067+0.107n_e+0.379\overline{ v_{ex}}^2
% \end{equation}
% \begin{equation}\nonumber 
%     -0.293\overline{ v_{ez}}^2 -0.072B_x^2-0.062B_zn_e+0.053B_z^2
% \end{equation}
\begin{align}\label{eqn22}
P_{e\parallel Lab} =
&-0.069+0.113n_e+0.370\overline{ v_{ex}}^2 -0.291\overline{ v_{ez}}^2 \nonumber 
    \\ &-0.078B_x^2+0.065B_z^2 -0.075B_zn_e
\end{align}
We readily notice that the SR-Lab closure in Equation \ref{eqn22} is not Galilean invariant since the terms proportional to $\{\overline{ v_{ex}}^2,\overline{ v_{ez}}^2\}$ are not Galilean invariant, e.g. $\overline{ v_{ez}'}^2 = (\overline{ v_{ez}}-\beta_z)^2\neq \overline{ v_{ez}} ^2$. The identification of these terms by SR are the result of a spurious correlation with the parallel electron pressure in the original (lab-frame) dataset. However, this correlation is broken when the dynamics is observed from different Galilean reference frames. Indeed, adding Galilean-augmented data in the SR closure inference removes these spurious terms as all Galilean reference frames must agree on the same closure. To embed Galilean symmetry, we double the training dataset size by including random Galilean boosts of each sampled point with $|\beta_j|\leq 0.4c$. With Galilean-boosted data, the following six-term model (SR-Boost) is inferred: 
% \begin{equation}\label{eqn23}
%     P_{e\parallel boost}=-0.068+0.098n_e-0.036B_x^2
% \end{equation}
% \begin{equation}\nonumber 
%     +0.192B_y^2-0.051B_zn_e+0.050B_zB_x
% \end{equation}
\begin{align}
    P_{e\parallel Boost} = & -0.071 +0.103n_e-0.040B_x^2
    \nonumber 
    \\ & +0.193B_y^2+0.062B_z^2-0.062B_zn_e,
    \label{eqn23}
\end{align}
which is readily observed to be Galilean invariant. A more detailed discussion of the inferred models and their physical interpretation will be left to future work. We note that for weaker Galilean boosts, for example $|\beta_j|\leq 0.1c$, the symmetry breaking terms are not removed from the closure inferred with augmented data; a sufficiently strong boost is required to break the spurious correlations between model residual and unwanted terms. 

Importantly, the SR-Boost closure shows significantly better generalization to unseen times 
$t > t_{max}$ compared to SR-Lab [Figure \ref{fig6}]. During the training window ($t < t_{max}$), both SR-Boost and SR-Lab are found to have similar model errors [Figure \ref{fig6}]; the removal of the spurious terms from SR-Boost results in only a small increase of the model error with respect to SR-Lab. However, the detrimental effects of the spurious terms on the accuracy of the closure models become most noticeable for $t>t_{max}$. Indeed, at these times we observe that the spurious correlations were not only frame-dependent, but also only occurred in a narrow range of time (during the training window). Elimination of these spurious terms through Galilean symmetry-embedding is crucial to enable a more robust and generalizable closure model.

In Figure \ref{fig6}, we further compare the obtained SR-Lab and SR-Boost closures with commonly used analytical pressure closure models for collisionless magnetic reconnection. These include the isothermal closure (where the pressure is linearly proportional to the density), the double adiabatic CGL closure model \cite{CGL1956}, and Le closure~\cite{Le2009}, which interpolates between isothermal and CGL closures depending on whether the plasma is in a passing or trapped state; we note that we are extending the Le closure outside of its regime of validity, $v_{the}\gg v_A$ ($v_{the}$ and $v_A$ are the electron thermal and Alfv\'en speed, respectively), which does not hold in our low beta pair plasma where $v_{the}/v_A=0.22$.  With only the original, lab-frame data the SR-Lab closure does not outperform CGL (the best performing analytical model for the simulated magnetic reconnection regime). However, including Galilean boosted data, the SR-Boost closure outperforms all other analytical closures, having the lowest model error at late times, beyond the training time horizon $t>t_{max}=250w_{pe}^{-1}$ [Figure \ref{fig6}]. At these late times the analytical models greatly underestimate $P_{e\parallel}$ leading to their large errors, and the improved performance of SR-Boost over SR-Lab is primarily due to its more accurate predictions upstream of the reconnecting current sheet, outside the plasmoid regions. We present in Appendix \ref{app8} a more detailed discussion about the error distributions of the pressure closure models used in Figure \ref{fig6}.

% At early times $t<t_{max}$ (during the training phase), however, the analytical closure models out-perform both SR models. This suggests that the SR library of terms is not expressive enough to simultaneously describe the pressure closure at both early and late times. We could attempt to expand the SR library of candidate terms to include higher more complex nonlinearities. Alternatively, neural networks, which are universal approximators, can be used to test for the existence of better functional forms. We therefore explore the use of neural networks in the next section. Neural networks provide a more expressive functional basis, potentially enabling more accurate modeling of both early and late time dynamics. We explore their use in the next section.

We note that at early times ($t<t_{max}$), corresponding to the training phase, the analytical closure models outperform both SR-based approaches. This indicates that the current SR library lacks the expressiveness needed to accurately capture the pressure closure across both early and late time regimes. One possible remedy is to enrich the SR candidate library by incorporating more complex nonlinear terms. Alternatively, neural networks (being universal function approximators \cite{Hornik1989}) offer a flexible framework to explore more expressive functional forms. Motivated by this, we turn to neural network-based closures in the next section.

\subsubsection*{\label{sec:level3:9}Neural Network Inferred Closure}

Neural networks (NNs), such as multilayer perceptrons, are able to model much more sophisticated functional forms than those in the simple library of catalog terms used in SR to infer the $P_{e\parallel}$ closure. This comes at the cost of reduced interpretability of the inferred model. Nevertheless, NNs can be used as tools to investigate the existence of more accurate closure models than those found by SR, by leveraging their ability to approximate more complex functional forms.

We therefore design NNs to approximate $P_{e\parallel}$ based on local and instantaneous values of lower-order fluid-moments and fields $\{n_e,\overline{v_{ex}},\overline{v_{ey}},\overline{v_{ez}}, E_x,E_y,E_z,B_x,B_y,B_z\}$ as inputs. Two NN models are trained using lab-frame-only data (which we refer to as NN-Lab) and Galilean-augmented data (NN-Boost). Training data is sampled and augmented using the procedure previously applied to SR. As before, the NNs are trained up to $t<t_{max}=250w_{pe}^{-1}$ and tested on their ability to generalize to later times; we explore the impact of varying $t_{max}$ in Appendix \ref{app9}. The NN architectures are composed of three hidden layers with 25 nodes per layer. They are trained in PyTorch~\cite{Paszke2019} for 2,000 epochs with an AdamW optimizer, learning rate of 0.001, and batch size of 10. These training parameters consistently lead to well-converged models.

The accuracy of the NN closure models (as measured by their mean absolute error) is shown in Figure~\ref{fig6}. For $t<t_{max}$, both NN models are observed to perform similarly to the analytical closure models, and outperform the SR pressure closure models. The improved NN model error at early times suggests the existence of other functional forms that better approximate the pressure closure model. These results can motivate a re-examination of the library of candidate terms used in SR to capture such improved behaviors in an interpretable framework.

For $t>t_{max}$, beyond the training horizon, we see a notable improvement in the performance of NN-Boost over NN-Lab [Figure \ref{fig6}]. These results reveal that NN-Boost generalizes better than NN-Lab, underscoring the benefits of symmetry-embedding via data augmentation in improving the generalization capabilities of ML models other than SR. However, despite their greater flexibility, the NN models do not outperform the SR-Boost closure in this regime. This indicates that the quadratic polynomial basis used in SR is already sufficient to capture the essential behavior of $P_{e\parallel}$ at late times. 

 These numerical experiments illustrate the importance of embedding fundamental physical symmetries in machine-learned-based fluid closure models from first-principles kinetic simulations. However, there are several important considerations that are worth discussing before applying these techniques to study magnetic reconnection, or other plasma phenomena more broadly. First, it is important to recognize that despite embedding fundamental physical symmetries, the validity of ML fluid closure models may be limited to the physical regimes in which they are trained. For example, ML closures trained on kinetic simulations of sub-relativistic magnetic reconnection, may not generalize to relativistic magnetic reconnection regimes, where fundamentally new behavior can arise, such as relativistic plasma effects and highly efficient nonthermal particle acceleration. These new effects may require fundamental structural changes to the closure model or to the reduced plasma framework itself (such as going from a pure plasma fluid framework to a hybrid kinetic-fluid plasma framework). Therefore, it is important to identify the range of physical regimes of interest at the onset, and define an appropriate reduced theoretical plasma framework accordingly. The kinetic simulation data used to train the ML closure models of such a framework should then span the physical regimes of interest -- with the aid of symmetry-guided data augmentation.

 Second, the ML pressure closure models explored in this work could potentially be more accurate by allowing for non-local dependencies (in space and in time) on low-order plasma moments and fields. Indeed, the inclusion of kinetic effects in fluid plasma models has been shown to require closure models that are inherently non-local \cite{Hammett1990,Held2004}. Such nonlocal relations can be captured using convolution neural networks \cite{Maulik2020, Ma2020, Bois2022, McgraeMenge2024,Huang2025}, for example. In addition, the development of ML closure models at a higher-order moment of the distribution function, such as the heat flux tensor for example, could also be fruitful. An ML closure model for the heat flux tensor (or higher order moment) may turn out to be simpler than that for the pressure tensor for a given problem, and may ultimately even facilitate gaining new insights into the underlying physics.

 Finally, to properly assess the performance of ML closures, it is important that these be implemented into a fluid (or hybrid) simulation framework, enabling self-consistent calculations with this model, and verifying if key observables of interest (such as the reconnection rate and energy partitioning, for example) are well captured. There has been recent progress in this direction \cite{Bois2022,Huang2022,Qin2023,Greif2023,Barbour2025,Carvalho2024,Joglekar2023,Huang2025}, but extending these approaches to magnetic reconnection and to higher-dimensional plasma phenomena remains a valuable direction for future work.

\section{\label{sec:discuss}Discussion}
In this work, we showed that data augmentation is an effective and flexible method to embed physical symmetries into machine-learned reduced plasma models trained on data from first-principle plasma simulations. This approach works across different types of machine learning models, physical problems, and physical symmetries. As a proof of concept, we showed that by Lorentz transforming PIC simulation data of collisionless magnetic reconnection we were able to embed Lorentz invariance into the SR inference of the relativistic two-fluid equations. Discovering reduced plasma models which abide by fundamental physical symmetries such as Lorentz invariance is crucial for their physical consistency and generalization performance. The act of embedding Lorentz invariance into the inference of the two-fluid equations resulted in more accurate coefficients and the elimination of unphysical, spurious model terms. We further showed that with a limited amount of original, simulation data combined with Lorentz-boosted data augmentation, we could discover more accurate models than with orders of magnitude more original simulation data. This has important implications for reducing the volume of data produced by state-of-the-art simulations, which is an increasingly pressing issue in the era of exascale computing \cite{Chang2023}.

We further demonstrated the importance of embedding symmetries into the inference of pressure closures for collisionless magnetic reconnection in the low Lorentz boost limit (Galilean boosts). We obtained a parallel pressure closure model with SR that was not only Galilean-invariant, but furthermore outperformed commonly used analytical closure models. To show the methodological generality of symmetry embedding with data augmentation we trained NN models to predict the parallel pressure with and without Galilean boosts. The symmetry-preserving models outperformed their original data counterparts. We envision further improvements to the development of ML fluid moment closures for magnetic reconnection by extending the library of SR terms or NN architectures to account for non-local effects \cite{McgraeMenge2024}. In addition, beyond fluid moment closures, the symmetry embedding techniques explored here can also be applied to the development of sub-grid models for anomalous resistivity and viscosity, which are important ingredients for theoretical models of magnetic reconnection at very large scales \cite{Selvi2023_2,Bugli2025,Moran2025}.

We note that rather than weakly embedding Galilean and Lorentz symmetries through data augmentation, we could have hard-constrained the symmetries by selecting inputs that were invariant by construction. In our example this would correspond to formulating a library of catalog terms comprised of Lorentz tensors. This is certainly a reasonable approach, however demands a careful method to construct the library of catalog terms that guarantees invariance. We aim to explore this approach in future work.

%Given decades of theoretical pursuit to understand the symmetries associated with plasma behavior in a multitude of environments, from fusion plasmas to astro-physical plasmas, it is an ever too often wasted opportunity that we discover reduced plasma models with machine learning algorithms without embedding these known physical laws. Data augmentation offers an approachable method of embedding these known physical symmetries, however complex or non-linear, and we envision its vital tool in ensuring the physical consistency of reduced plasma modeling of multiscale systems at the forefront of future plasma research.

In summary, we have shown that symmetry‑guided data augmentation provides a powerful and versatile approach for embedding physical symmetries into machine‑learned reduced models of plasma dynamics constructed from the data of first‑principles kinetic simulations. This approach is agnostic to the learning framework, integrating seamlessly with techniques that range from sparse regression to deep neural networks. Our results help lay the foundation for the development of more accurate, physically‑consistent, and data-efficient reduced plasma models, which are key to enabling computationally efficient multiscale simulations of complex plasma systems.

\begin{acknowledgments}
The authors would like to thank M. Almanza, N. Barbour, D. Carvalho, W. Dorland, G. Guttormsen, N. Loureiro, G. Morales, I. Pusztai, A. Vanthieghem, and A. Velberg for helpful discussions. The authors acknowledge the OSIRIS Consortium, consisting of UCLA, University of Michigan, and IST (Portugal) for the use of the OSIRIS 4.0 framework. This work was supported by the National Science Foundation Grants No. PHY-2108087 and No. PHY-2108089. F.F. acknowledges support from the European Research Council (ERC-2021-CoG Grant XPACE No. 101045172). Simulations were run on Cori and Perlmutter (NERSC) through ALCC awards. 
\end{acknowledgments}

\appendix

\section{\label{app1}Simulation Details}
Here we detail the physical and numerical parameters of our particle-in-cell simulation of collisionless magnetic reconnection. We construct a force-free current sheet with the following magnetic field profiles:
% \begin{equation}
%     B_x(y) = B_{reconn}[\tanh(\frac{y+\frac{L_y}{4}}{\lambda})-\tanh(\frac{y-\frac{L_y}{4}}{\lambda})-1]
% \end{equation}
% \begin{equation}
%     B_z(y) = \sqrt{B_{reconn}^2+B_{guide}^2-B_x(y)}
% \end{equation}
\begin{align}
    B_x(y) &= B_{reconn}\left[ \tanh \left( \frac{y+\frac{L_y}{4}}{\lambda} \right) -
    \tanh\left( \frac{y-\frac{L_y}{4}}{\lambda} \right)-1 \right]
    \\ B_z(y) &= \sqrt{B_{reconn}^2+B_{guide}^2-B_x(y)}
\end{align}
This results in two current sheets spatially separated in $\bm{\hat{y}}$ with out-of-the-plane, $\bm{\hat{z}}$ currents in opposite directions. It is periodic in both $(x,y)$. These magnetic field profiles satisfy the force-free equilibrium condition $\bm{J}\times\bm{B}=0$. There is no pressure gradient in the initialization. To perturb the system in an asymmetric manner we initialize: 
\begin{equation}
    B_y(x) = \delta B \left[ \sin \left( \frac{2\pi x}{L_x} \right)
    +\frac{1}{4}\exp \left( {-\left(\frac{x-\frac{L_x}{4}}{(\frac{L_x}{8})} \right)^2} \right) \right].
\end{equation}
In our system $B_{reconn}=0.5$, $B_{guide}=0.2$, $\delta B=0.01$, $L_x=48d_e$, $L_y=64d_e$, and $\lambda=1d_e$. This is the medium guide field regime as $\frac{B_{guide}}{B_{reconn}}=0.4$. From the magnetic fields we calculate the self-consistent current density. We partition the initial density and momentum equally between electrons and positrons ($n_e=n_p=n_0=1$, $\bm{v_e}=-\frac{\bm{J}}{2}$, $\bm{v_p}=\frac{\bm{J}}{2}$). We choose $\beta_e=\beta_p=\beta_0=0.1$, where $\beta_s\equiv p_s / (B^2/8\pi)$, which sets the electron and positron thermal velocities $v_{th}=\sqrt{\frac{\beta_0}{8}}\approx 0.112c$. For Section~\ref{sec:level2:1}, the numerical PIC parameters used were 1600 particles per cell, linear particle shapes, grid spacing $\Delta x=\Delta y = 0.1d_e$, and timestep $\Delta t=0.0667w_{pe}^{-1}$. The simulation is non-relativistic as the initial ratio of thermal to magnetic energy $\sim \frac{v_{th}^2}{B_{reconn}^2}\approx0.05$. We end the simulation at $t_{max}=10^3w_{pe}^{-1}$ when the reconnection rate has saturated. The data used for the pressure closure model discovery in Section~\ref{sec:level2:2} was generated using identical physical parameters, but used 625 particles-per-cell and quadratic particle shapes.

\section{\label{app3}Candidate Terms for Two-Fluid Equation Re-Discovery}

Here we provide details for the sparse regression inference of the two-fluid moment equations from magnetic reconnection data. For each equation inference we provided sparse regression with the possible catalog terms to describe the target (dynamical) variable. The choice of these catalog terms is guided by the physical intuition of the problem; in this case we know the true equations and thus build catalog term libraries which include the true model terms, as well as other reasonable model terms which may throw the sparse regression algorithm off from the true equation. The catalog terms are the building block catalog terms terms listed below and polynomial combinations of the building block terms up to order 2.  
\begin{center}
\begin{tabular}{|p{18.5mm}|p{19.5mm}|p{35mm}| } 
\hline
\textbf{Equation} & \textbf{Target} \newline \textbf{Variable} & \textbf{Building Block} \newline \textbf{Catalog Terms} \\
\hline
Continuity & $\partial_t n_e$& $1,n_e,n_e\overline{v_{ex}},n_e\overline{v_{ey}},$\newline $n_e\overline{v_{ez}},\partial_x[n_e\overline{v_{ex}}],$ \newline $\partial_y[n_e\overline{v_{ey}}],E_x,E_y,$\newline $E_z,B_x,B_y,B_z$\\ \hline 
$\bm{\hat{x}}$\newline Momentum & $\partial_t[n_e\overline{p_{ex}}]$ & $1,n_e,n_e\overline{v_{ex}},n_e\overline{v_{ey}},$\newline $n_e\overline{v_{ez}},n_e\overline{v_{ex}p_{ex}},$\newline $n_e\overline{v_{ex}p_{ey}},n_e\overline{v_{ex}p_{ez}},$\newline $\partial_x[n_e\overline{v_{ex}p_{ex}} ],$\newline $\partial_y[n_e\overline{v_{ex}p_{ey}}],$\newline $E_x,E_y,E_z,B_x,B_y,B_z$ \\ \hline 

$\bm{\hat{y}}$\newline Momentum 
& $\partial_t[n_e\overline{p_{ey}}]$ & $1,n_e,n_e\overline{v_{ex}},n_e\overline{v_{ey}},$\newline $n_e\overline{v_{ez}},n_e\overline{v_{ey}p_{ex}},$\newline $n_e\overline{v_{ey}p_{ey}},n_e\overline{v_{ey}p_{ez}},$\newline $\partial_x[n_e\overline{v_{ey}p_{ex}} ],$\newline $\partial_y[n_e\overline{v_{ey}p_{ey}}],$\newline $E_x,E_y,E_z,B_x,B_y,B_z$ \\

\hline
$\bm{\hat{z}}$\newline Momentum 
& $\partial_t[n_e\overline{p_{ez}}]$ & $1,n_e,n_e\overline{v_{ex}},n_e\overline{v_{ey}},$\newline $n_e\overline{v_{ez}},n_e\overline{v_{ez}p_{ex}},$\newline $n_e\overline{v_{ez}p_{ey}},n_e\overline{v_{ez}p_{ez}},$\newline $\partial_x[n_e\overline{v_{ez}p_{ex}} ],$\newline $\partial_y[n_e\overline{v_{ez}p_{ey}}],$\newline $E_x,E_y,E_z,B_x,B_y,B_z$ \\ 
\hline

$\bm{\hat{x}}\bm{\hat{x}}$\newline Energy 
& $\partial_t[n_e\overline{p_{ex}p_{ex}}]$ & $1,n_e,n_e\overline{p_{ex}},n_e\overline{p_{ey}},$\newline $n_e\overline{p_{ez}},n_e\overline{v_{ex}p_{ex}},$\newline $n_e\overline{v_{ex}p_{ey}},n_e\overline{v_{ex}p_{ez}},$\newline $n_e\overline{v_{ex}p_{ex}p_{ex}} ,$\newline $n_e\overline{v_{ey}p_{ex}p_{ex}},$\newline $n_e\overline{v_{ez}p_{ex}p_{ex}},$\newline $\partial_x[n_e\overline{v_{ex}p_{ex}p_{ex}} ],$\newline $\partial_y[n_e\overline{v_{ey}p_{ex}p_{ex}}],$\newline $E_x,E_y,E_z,B_x,B_y,B_z$ \\ 
\hline
$\bm{\hat{y}}\bm{\hat{y}}$\newline Energy 
& $\partial_t[n_e\overline{p_{ey}p_{ey}}]$ & $1,n_e,n_e\overline{p_{ex}},n_e\overline{p_{ey}},$\newline $n_e\overline{p_{ez}},n_e\overline{v_{ey}p_{ex}},$\newline $n_e\overline{v_{ey}p_{ey}},n_e\overline{v_{ey}p_{ez}},$\newline $n_e\overline{v_{ex}p_{ey}p_{ey}} ,$\newline $n_e\overline{v_{ey}p_{ey}p_{ey}},$\newline $n_e\overline{v_{ez}p_{ey}p_{ey}},$\newline $\partial_x[n_e\overline{v_{ex}p_{ey}p_{ey}} ],$\newline $\partial_y[n_e\overline{v_{ey}p_{ey}p_{ey}}],$\newline $E_x,E_y,E_z,B_x,B_y,B_z$ \\ 
\hline 
$\bm{\hat{z}}\bm{\hat{z}}$\newline Energy 
& $\partial_t[n_e\overline{p_{ez}p_{ez}}]$ & $1,n_e,n_e\overline{p_{ex}},n_e\overline{p_{ey}},$\newline $n_e\overline{p_{ez}},n_e\overline{v_{ez}p_{ex}},$\newline $n_e\overline{v_{ez}p_{ey}},n_e\overline{v_{ez}p_{ez}},$\newline $n_e\overline{v_{ex}p_{ez}p_{ez}} ,$\newline $n_e\overline{v_{ey}p_{ez}p_{ez}},$\newline $n_e\overline{v_{ez}p_{ez}p_{ez}},$\newline $\partial_x[n_e\overline{v_{ex}p_{ez}p_{ez}} ],$\newline $\partial_y[n_e\overline{v_{ey}p_{ez}p_{ez}}],$\newline $E_x,E_y,E_z,B_x,B_y,B_z$ \\
\hline 
\end{tabular}
\end{center}

\begin{center}
\begin{tabular}{|p{18.5mm}|p{19.5mm}|p{34mm}| } 
\hline

$\bm{\hat{x}}\bm{\hat{y}}$\newline Energy 
& $\partial_t[n_e\overline{p_{ex}p_{ey}}]$ & $1,n_e,n_e\overline{p_{ex}},n_e\overline{p_{ey}},$\newline $n_e\overline{p_{ez}},n_e\overline{v_{ex}p_{ex}},$\newline $n_e\overline{v_{ey}p_{ey}},n_e\overline{v_{ez}p_{ez}},$\newline 
$n_e\overline{v_{ex}p_{ey}},n_e\overline{v_{ex}p_{ez}},$\newline 
$n_e\overline{v_{ey}p_{ez}},$
\newline $n_e\overline{v_{ex}p_{ex}p_{ey}} ,$\newline $n_e\overline{v_{ey}p_{ex}p_{ey}},$\newline $n_e\overline{v_{ez}p_{ex}p_{ey}},$\newline $\partial_x[n_e\overline{v_{ex}p_{ex}p_{ey}} ],$\newline $\partial_y[n_e\overline{v_{ey}p_{ex}p_{ey}}],$\newline $E_x,E_y,E_z,B_x,B_y,B_z$ \\ 
\hline

$\bm{\hat{x}}\bm{\hat{z}}$\newline Energy 
& $\partial_t[n_e\overline{p_{ex}p_{ez}}]$ & $1,n_e,n_e\overline{p_{ex}},n_e\overline{p_{ey}},$\newline $n_e\overline{p_{ez}},n_e\overline{v_{ex}p_{ex}},$\newline $n_e\overline{v_{ey}p_{ey}},n_e\overline{v_{ez}p_{ez}},$\newline 
$n_e\overline{v_{ex}p_{ey}},n_e\overline{v_{ex}p_{ez}},$\newline 
$n_e\overline{v_{ey}p_{ez}},$
\newline $n_e\overline{v_{ex}p_{ex}p_{ez}} ,$\newline $n_e\overline{v_{ey}p_{ex}p_{ez}},$\newline $n_e\overline{v_{ez}p_{ex}p_{ez}},$\newline $\partial_x[n_e\overline{v_{ex}p_{ex}p_{ez}} ],$\newline $\partial_y[n_e\overline{v_{ey}p_{ex}p_{ez}}],$\newline $E_x,E_y,E_z,B_x,B_y,B_z$ \\ 

\hline
$\bm{\hat{y}}\bm{\hat{z}}$\newline Energy 
& $\partial_t[n_e\overline{p_{ey}p_{ez}}]$ & $1,n_e,n_e\overline{p_{ex}},n_e\overline{p_{ey}},$\newline $n_e\overline{p_{ez}},n_e\overline{v_{ex}p_{ex}},$\newline $n_e\overline{v_{ey}p_{ey}},n_e\overline{v_{ez}p_{ez}},$\newline 
$n_e\overline{v_{ex}p_{ey}},n_e\overline{v_{ex}p_{ez}},$\newline 
$n_e\overline{v_{ey}p_{ez}},$
\newline $n_e\overline{v_{ex}p_{ey}p_{ez}} ,$\newline $n_e\overline{v_{ey}p_{ey}p_{ez}},$\newline $n_e\overline{v_{ez}p_{ey}p_{ez}},$\newline $\partial_x[n_e\overline{v_{ex}p_{ey}p_{ez}} ],$\newline $\partial_y[n_e\overline{v_{ey}p_{ey}p_{ez}}],$\newline $E_x,E_y,E_z,B_x,B_y,B_z$ \\ 
\hline
\end{tabular}
\end{center}
\section{\label{app2}Data Sampling Details}
Here we detail the data sampling procedure to extract measurements for the sparse regression recovery of the two-fluid equations. First, to remove some of the data overhead we sample our data from windows (clusters) of neighboring timesteps. Beginning at timestep $3,500$ (which corresponds to a physical time $t=233.33w_{pe}^{-1}$) we sample our measurements from a $30$ timestep ($2w_{pe}^{-1}$) window. We then skip $800$ timesteps ($53.33w_{pe}^{-1}$) until the next $30$ timestep window. The process is repeated until timestep $6,020$ ($401.33w_{pe}^{-1}$). This results in four time-windows of consecutive timesteps where we can sample measurements for the sparse regression equation inference. These time-windows fall during the non-linear stage of MR when complex, non-linear, and chaotic dynamics are occurring. To ensure that the measured points reside in interesting regions in space-time where the target variables are dynamically balanced by the interplay of model terms, for each time-window we calculate the $70$th percentile value of the absolute target variables and only select measurements above that threshold. The results are weakly sensitive to the percentile value above $50\%$. A total of $15^3=3,375$ measurements are randomly selected - satisfying the threshold condition - from the time-windows. Selecting a large amount of measurements from the simulations, as done here, helps reduce the variance of model coefficient error as observed in Figure \ref{fig2}. We then integrate these measurements over small space-time cubes of size $\{5\Delta x,5\Delta y,5\Delta t\}$ (the integral formulation) and discover equations in their integral form. The data sampling scheme is different for closure discovery and described in the text body [Section \ref{sec:level2:2}].

\section{\label{app5}Table of Inferred Coefficients}

Here we provide the two-fluid model equations inferred by sparse regression with and without Lorentz-boosted data transformations in Section \ref{sec:level2:1}. The Lorentz boosts are conducted with a maximum Lorentz boost of $\gamma_{\bm{\beta_{max}}}=1.5$. The error bars correspond to $\pm$ one standard deviation, and the statistics come from training each model five times with different realizations of the random data sampling. We compare the Lab-Frame and Boosted Data Equations with the analytically derived True Equations.

\begin{center}
\begin{tabular}{|p{22mm}|p{51mm}|} 
\hline \multicolumn{2}{| c |}{\textbf{Continuity Equation}}\\
\hline True \newline Equation & $\partial_t n_e=$\newline$-\partial_x[n_e\overline{ v_{ex}}]$\newline $
-\partial_y[n_e\overline{ v_{ey}}]$ \\ \hline Lab-Frame Data \newline Inferred \newline Equation & $\partial_t n_e=$\newline $-(1.0262\pm 0.0028)\partial_x[n_e\overline{ v_{ex}}]$ \newline $
-(1.0317\pm 0.0022)\partial_y[n_e\overline{ v_{ey}}]$\\ \hline 
Boosted Data \newline Inferred \newline Equation & $\partial_t n_e=$\newline $-(1.0006\pm 0.0007)\partial_x[n_e\overline{ v_{ex}}]$ \newline $
-(1.0005\pm 0.0001)\partial_y[n_e\overline{ v_{ey}}]$
\\ \hline 

\end{tabular}
\end{center}

\begin{center}
\begin{tabular}{|p{17mm}|p{56mm}|} 
\hline \multicolumn{2}{| c |}{\textbf{$\bm{\hat{x}}$ Momentum Equation}}\\
\hline True \newline Equation & $\partial_t [n_e\overline{ p_{ex}}]=$\newline$-\partial_x[n_e\overline{ v_{ex}p_{ex}}]$\newline $
-\partial_y[n_e\overline{ v_{ex}p_{ey}}]$\newline$-n_eE_x$\newline$-n_e\overline{ v_{ey}} B_z$\newline $+n_e\overline{ v_{ez}} B_y$ \\ \hline Lab-Frame \newline Data \newline Inferred \newline Equation & $\partial_t [n_e\overline{ p_{ex}}]=$\newline$-(1.0189\pm0.0008)\partial_x[n_e\overline{ v_{ex}p_{ex}}]$\newline $
-(1.0311\pm0.0014)\partial_y[n_e\overline{ v_{ex}p_{ey}}]$\newline$-(1.0071\pm0.0016)n_eE_x$\newline$-(1.0077\pm0.0018)n_e\overline{ v_{ey}} B_z$\newline $+(1.0107\pm0.0016)n_e\overline{ v_{ez}} B_y$ \\ \hline 
Boosted Data \newline Inferred \newline Equation & $\partial_t [n_e\overline{ p_{ex}}]=$\newline$-(1.0002\pm0.0005)\partial_x[n_e\overline{ v_{ex}p_{ex}}]$\newline $
-(1.0016\pm0.0003)\partial_y[n_e\overline{ v_{ex}p_{ey}}]$\newline$-(0.9918\pm0.0051)n_eE_x$\newline$-(0.9917\pm0.0050)n_e\overline{ v_{ey}} B_z$\newline $+(0.9931\pm0.0040)n_e\overline{ v_{ez}} B_y$
\\ \hline 

\end{tabular}
\end{center}

\begin{center}
\begin{tabular}{|p{17mm}|p{56mm}|} 
\hline \multicolumn{2}{| c |}{\textbf{$\bm{\hat{y}}$ Momentum Equation}}\\
\hline True \newline Equation & $\partial_t [n_e\overline{ p_{ey}}]=$\newline$-\partial_x[n_e\overline{ v_{ey}p_{ex}}]$\newline $
-\partial_y[n_e\overline{ v_{ey}p_{ey}}]$\newline$-n_eE_y$\newline$-n_e\overline{ v_{ez}} B_x$\newline $+n_e\overline{ v_{ex}} B_z$ \\ \hline Lab-Frame \newline Data \newline Inferred \newline Equation & $\partial_t [n_e\overline{ p_{ey}}]=$\newline$-(1.0211\pm0.0022)\partial_x[n_e\overline{ v_{ey}p_{ex}}]$\newline $
-(1.0245\pm0.0017)\partial_y[n_e\overline{ v_{ey}p_{ey}}]$\newline$-(1.0079\pm0.0023)n_eE_y$\newline$-(1.0081\pm0.0023)n_e\overline{ v_{ez}} B_x$\newline $+(1.0080\pm0.0021)n_e\overline{ v_{ex}} B_z$ \\ \hline 
Boosted Data \newline Inferred \newline Equation & $\partial_t [n_e\overline{ p_{ey}}]=$\newline$-(1.0012\pm0.0006)\partial_x[n_e\overline{ v_{ey}p_{ex}}]$\newline $
-(1.0010\pm0.0002)\partial_y[n_e\overline{ v_{ey}p_{ey}}]$\newline$-(0.9916\pm0.0018)n_eE_y$\newline$-(0.9925\pm0.0011)n_e\overline{ v_{ez}} B_x$\newline $+(0.9916\pm0.0018)n_e\overline{ v_{ex}} B_z$ \\ \hline 

\end{tabular}
\end{center}

\begin{center}
\begin{tabular}{|p{17mm}|p{56mm}|} 
\hline \multicolumn{2}{| c |}{\textbf{$\bm{\hat{z}}$ Momentum Equation}}\\
\hline True \newline Equation & $\partial_t [n_e\overline{ p_{ez}}]=$\newline$-\partial_x[n_e\overline{ v_{ez}p_{ex}}]$\newline $
-\partial_y[n_e\overline{ v_{ez}p_{ey}}]$\newline$-n_eE_z$\newline$-n_e\overline{ v_{ex}} B_y$\newline $+n_e\overline{ v_{ey}} B_x$ \\ \hline Lab-Frame \newline Data \newline Inferred \newline Equation & $\partial_t [n_e\overline{ p_{ez}}]=$\newline$-(1.0256\pm0.0042)\partial_x[n_e\overline{ v_{ez}p_{ex}}]$\newline $
-(1.0322\pm0.0026)\partial_y[n_e\overline{ v_{ez}p_{ey}}]$\newline$-(1.0207\pm0.0030)n_eE_z$\newline$-(1.0187\pm0.0019)n_e\overline{ v_{ex}} B_y$\newline $+(1.0173\pm0.0029)n_e\overline{ v_{ey}} B_x$ \\ \hline 
Boosted Data \newline Inferred \newline Equation & $\partial_t [n_e\overline{ p_{ez}}]=$\newline$-(1.0007\pm0.0005)\partial_x[n_e\overline{ v_{ez}p_{ex}}]$\newline $
-(1.0003\pm0.0002)\partial_y[n_e\overline{ v_{ez}p_{ey}}]$\newline$-(0.9942\pm0.0029)n_eE_z$\newline$-(0.9942\pm0.0029)n_e\overline{ v_{ex}} B_y$\newline $+(0.9942\pm0.0029)n_e\overline{ v_{ey}} B_x$ \\ \hline 

\end{tabular}
\end{center}

\begin{center}
\begin{tabular}{|p{17mm}|p{56mm}|} 
\hline \multicolumn{2}{| c |}{\textbf{$\bm{\hat{x}}\bm{\hat{x}}$ Energy Equation}}\\
\hline True \newline Equation & $\partial_t [n_e\overline{ p_{ex}p_{ex}}]=$\newline$-\partial_x[n_e\overline{ v_{ex}p_{ex}p_{ex}}]$\newline $
-\partial_y[n_e\overline{ v_{ey}p_{ex}p_{ex}}]$\newline$-2n_e\overline{ p_{ex}} E_x$\newline$-2n_e\overline{ v_{ex}p_{ey}} B_z$\newline $+2n_e\overline{ v_{ex}p_{ez}} B_y$ \\ \hline Lab-Frame \newline Data \newline Inferred \newline Equation & $\partial_t [n_e\overline{ p_{ex}p_{ex}}]=$\newline$-(1.0219\pm0.0020)\partial_x[n_e\overline{ v_{ex}p_{ex}p_{ex}}]$\newline $
-(1.0360\pm0.0021)\partial_y[n_e\overline{ v_{ey}p_{ex}p_{ex}}]$\newline$-(2.0230\pm0.0086)n_e\overline{ p_{ex}} E_x$\newline$-(2.0230\pm0.0051)n_e\overline{ v_{ex}p_{ey}} B_z$\newline $+(2.0255\pm0.0049)n_e\overline{ v_{ex}p_{ez}} B_y$ \\ \hline 
Boosted \newline Data \newline Inferred \newline Equation & $\partial_t [n_e\overline{ p_{ex}p_{ex}}]=$\newline$-(1.0010\pm0.0004)\partial_x[n_e\overline{ v_{ex}p_{ex}p_{ex}}]$\newline $
-(1.0095\pm0.0016)\partial_y[n_e\overline{ v_{ey}p_{ex}p_{ex}}]$\newline$-(1.9881\pm0.0057)n_e\overline{ p_{ex}} E_x$\newline$-(1.9871\pm0.0052)n_e\overline{ v_{ex}p_{ey}} B_z$\newline $+(1.9959\pm0.0074)n_e\overline{ v_{ex}p_{ez}} B_y$ \\ \hline 

\end{tabular}
\end{center}

\begin{center}
\begin{tabular}{|p{17mm}|p{56mm}|} 
\hline \multicolumn{2}{| c |}{\textbf{$\bm{\hat{y}}\bm{\hat{y}}$ Energy Equation}}\\
\hline True \newline Equation & $\partial_t [n_e\overline{ p_{ey}p_{ey}}]=$\newline$-\partial_x[n_e\overline{ v_{ex}p_{ey}p_{ey}}]$\newline $
-\partial_y[n_e\overline{ v_{ey}p_{ey}p_{ey}}]$\newline$-2n_e\overline{ p_{ey}} E_y$\newline$-2n_e\overline{ v_{ey}p_{ez}} B_x$\newline $+2n_e\overline{ v_{ey}p_{ex}} B_z$ \\ \hline Lab-Frame \newline Data \newline Inferred \newline Equation & $\partial_t [n_e\overline{ p_{ey}p_{ey}}]=$\newline$-(1.0278\pm0.0036)\partial_x[n_e\overline{ v_{ex}p_{ey}p_{ey}}]$\newline $
-(1.0299\pm0.0023)\partial_y[n_e\overline{ v_{ey}p_{ey}p_{ey}}]$\newline$-(2.0263\pm0.0093)n_e\overline{ p_{ey}} E_y$\newline$-(2.0250\pm0.0069)n_e\overline{ v_{ey}p_{ez}} B_x$\newline $+(2.0247\pm0.0059)n_e\overline{ v_{ey}p_{ex}} B_z$ \\ \hline 
Boosted \newline Data \newline Inferred \newline Equation & $\partial_t [n_e\overline{ p_{ey}p_{ey}}]=$\newline$-(1.0062\pm0.0027)\partial_x[n_e\overline{ v_{ex}p_{ey}p_{ey}}]$\newline $
-(1.0016\pm0.0004)\partial_y[n_e\overline{ v_{ey}p_{ey}p_{ey}}]$\newline$-(1.9965\pm0.0087)n_e\overline{ p_{ey}} E_y$\newline$-(1.9979\pm0.0062)n_e\overline{ v_{ey}p_{ez}} B_x$\newline $+(1.9971\pm0.0084)n_e\overline{ v_{ey}p_{ex}} B_z$ \\ \hline 

\end{tabular}
\end{center}

\begin{center}
\begin{tabular}{|p{17mm}|p{56mm}|} 
\hline \multicolumn{2}{| c |}{\textbf{$\bm{\hat{z}}\bm{\hat{z}}$ Energy Equation}}\\
\hline True \newline Equation & $\partial_t [n_e\overline{ p_{ez}p_{ez}}]=$\newline$-\partial_x[n_e\overline{ v_{ex}p_{ez}p_{ez}}]$\newline $
-\partial_y[n_e\overline{ v_{ey}p_{ez}p_{ez}}]$\newline$-2n_e\overline{ p_{ez}} E_z$\newline$-2n_e\overline{ v_{ez}p_{ex}} B_y$\newline $+2n_e\overline{ v_{ez}p_{ey}} B_x$ \\ \hline Lab-Frame \newline Data \newline Inferred \newline Equation & $\partial_t [n_e\overline{ p_{ez}p_{ez}}]=$\newline$-(1.0287\pm0.0048)\partial_x[n_e\overline{ v_{ex}p_{ez}p_{ez}}]$\newline $
-(1.0361\pm0.0054)\partial_y[n_e\overline{ v_{ey}p_{ez}p_{ez}}]$\newline$-(2.0440\pm0.0073)n_e\overline{ p_{ez}} E_z$\newline$-(2.0290\pm0.0075)n_e\overline{ v_{ez}p_{ex}} B_y$\newline $+(2.0259\pm0.0082)n_e\overline{ v_{ez}p_{ey}} B_x$ \\ \hline 
Boosted \newline Data \newline Inferred \newline Equation & $\partial_t [n_e\overline{ p_{ez}p_{ez}}]=$\newline$-(1.0006\pm0.0004)\partial_x[n_e\overline{ v_{ex}p_{ez}p_{ez}}]$\newline $
-(1.0004\pm0.0003)\partial_y[n_e\overline{ v_{ey}p_{ez}p_{ez}}]$\newline$-(1.9965\pm0.0040)n_e\overline{ p_{ez}} E_z$\newline$-(1.9963\pm0.0042)n_e\overline{ v_{ez}p_{ex}} B_y$\newline $+(1.9961\pm0.0041)n_e\overline{ v_{ez}p_{ey}} B_x$ \\ \hline 

\end{tabular}
\end{center}

\begin{center}
\begin{tabular}{|p{17mm}|p{57mm}|} 
\hline \multicolumn{2}{| c |}{\textbf{$\bm{\hat{x}}\bm{\hat{y}}$ Energy Equation}}\\
\hline True \newline Equation & $\partial_t [n_e\overline{ p_{ex}p_{ey}}]=$\newline$-\partial_x[n_e\overline{ v_{ex}p_{ex}p_{ey}}]$\newline $
-\partial_y[n_e\overline{ v_{ey}p_{ex}p_{ey}}]$\newline$-n_e\overline{ p_{ex}} E_y$\newline$-n_e\overline{ p_{ey}} E_x$\newline$-n_e\overline{ v_{ex}p_{ez}} B_x$\newline $-n_e\overline{ v_{ey}p_{ey}} B_z$\newline $+n_e\overline{ v_{ey}p_{ez}} B_y$\newline $+n_e\overline{ v_{ex}p_{ex}} B_z$ \\ \hline Lab-Frame \newline Data \newline Inferred \newline Equation & $\partial_t [n_e\overline{ p_{ex}p_{ey}}]=$\newline$-(1.0324\pm0.0033)\partial_x[n_e\overline{ v_{ex}p_{ex}p_{ey}}]$\newline $
-(1.0418\pm0.0023)\partial_y[n_e\overline{ v_{ey}p_{ex}p_{ey}}]$\newline$-(1.0184\pm0.0049)n_e\overline{ p_{ex}} E_y$\newline$-(1.0171\pm0.0038)n_e\overline{ p_{ey}} E_x$\newline$-(1.0193\pm0.0031)n_e\overline{ v_{ex}p_{ez}} B_x$\newline $-(1.0173\pm0.0029)n_e\overline{ v_{ey}p_{ey}} B_z$\newline $+(1.0202\pm0.0037)n_e\overline{ v_{ey}p_{ez}} B_y$\newline $+(1.0177\pm0.0030)n_e\overline{ v_{ex}p_{ex}} B_z$ \\ \hline 
Boosted \newline Data \newline Inferred \newline Equation & $\partial_t [n_e\overline{ p_{ex}p_{ey}}]=$\newline$-(0.9999\pm0.0003)\partial_x[n_e\overline{ v_{ex}p_{ex}p_{ey}}]$\newline $
-(1.0009\pm0.0003)\partial_y[n_e\overline{ v_{ey}p_{ex}p_{ey}}]$\newline$-(0.9882\pm0.0017)n_e\overline{ p_{ex}} E_y$\newline$-(0.9913\pm0.0021)n_e\overline{ p_{ey}} E_x$\newline$-(0.9887\pm0.0016)n_e\overline{ v_{ex}p_{ez}} B_x$\newline $-(0.9912\pm0.0021)n_e\overline{ v_{ey}p_{ey}} B_z$\newline $+(0.9917\pm0.0019)n_e\overline{ v_{ey}p_{ez}} B_y$\newline $+(0.9882\pm0.0017)n_e\overline{ v_{ex}p_{ex}} B_z$ \\ \hline 

\end{tabular}
\end{center}

\begin{center}
\begin{tabular}{|p{17mm}|p{56mm}|} 
\hline \multicolumn{2}{| c |}{\textbf{$\bm{\hat{x}}\bm{\hat{z}}$ Energy Equation}}\\
\hline True \newline Equation & $\partial_t [n_e\overline{ p_{ex}p_{ez}}]=$\newline$-\partial_x[n_e\overline{ v_{ex}p_{ex}p_{ez}}]$\newline $
-\partial_y[n_e\overline{ v_{ey}p_{ex}p_{ez}}]$\newline$-n_e\overline{ p_{ex}} E_z$\newline$-n_e\overline{ p_{ez}} E_x$\newline$-n_e\overline{ v_{ey}p_{ez}} B_z$\newline $-n_e\overline{ v_{ex}p_{ex}} B_y$\newline $+n_e\overline{ v_{ex}p_{ey}} B_x$\newline $+n_e\overline{ v_{ez}p_{ez}} B_y$ \\ \hline Lab-Frame \newline Data \newline Inferred \newline Equation & $\partial_t [n_e\overline{ p_{ex}p_{ez}}]=$\newline$-(1.0280\pm0.0024)\partial_x[n_e\overline{ v_{ex}p_{ex}p_{ez}}]$\newline $
-(1.0429\pm0.0033)\partial_y[n_e\overline{ v_{ey}p_{ex}p_{ez}}]$\newline$-(1.0246\pm0.0025)n_e\overline{ p_{ex}} E_z$\newline$-(1.0189\pm0.0018)n_e\overline{ p_{ez}} E_x$\newline$-(1.0150\pm0.0012)n_e\overline{ v_{ey}p_{ez}} B_z$\newline $-(1.0181\pm0.0011)n_e\overline{ v_{ex}p_{ex}} B_y$\newline $+(1.0175\pm0.0026)n_e\overline{ v_{ex}p_{ey}} B_x$\newline $+(1.0172\pm0.0012)n_e\overline{ v_{ez}p_{ez}} B_y$ \\ \hline 
Boosted \newline Data \newline Inferred \newline Equation & $\partial_t [n_e\overline{ p_{ex}p_{ez}}]=$\newline$-(1.0004\pm0.0005)\partial_x[n_e\overline{ v_{ex}p_{ex}p_{ez}}]$\newline $
-(1.0015\pm0.0002)\partial_y[n_e\overline{ v_{ey}p_{ex}p_{ez}}]$\newline$-(0.9952\pm0.0078)n_e\overline{ p_{ex}} E_z$\newline$-(0.9960\pm0.0046)n_e\overline{ p_{ez}} E_x$\newline$-(0.9958\pm0.0045)n_e\overline{ v_{ey}p_{ez}} B_z$\newline $-(0.9951\pm0.0078)n_e\overline{ v_{ex}p_{ex}} B_y$\newline $+(0.9951\pm0.0076)n_e\overline{ v_{ex}p_{ey}} B_x$\newline $+(0.9950\pm0.0048)n_e\overline{ v_{ez}p_{ez}} B_y$ \\ \hline 

\end{tabular}
\end{center}

\begin{center}
\begin{tabular}{|p{17mm}|p{56mm}|} 
\hline \multicolumn{2}{| c |}{\textbf{$\bm{\hat{y}}\bm{\hat{z}}$ Energy Equation}}\\
\hline True \newline Equation & $\partial_t [n_e\overline{ p_{ey}p_{ez}}]=$\newline$-\partial_x[n_e\overline{ v_{ex}p_{ey}p_{ez}}]$\newline $
-\partial_y[n_e\overline{ v_{ey}p_{ey}p_{ez}}]$\newline$-n_e\overline{ p_{ey}} E_z$\newline$-n_e\overline{ p_{ez}} E_y$\newline$-n_e\overline{ v_{ex}p_{ey}} B_y$\newline $-n_e\overline{ v_{ez}p_{ez}} B_x$\newline $+n_e\overline{ v_{ex}p_{ez}} B_z$\newline $+n_e\overline{ v_{ey}p_{ey}} B_x$ \\ \hline Lab-Frame \newline Data \newline Inferred \newline Equation & $\partial_t [n_e\overline{ p_{ey}p_{ez}}]=$\newline$-(1.0252\pm0.0040)\partial_x[n_e\overline{ v_{ex}p_{ey}p_{ez}}]$\newline $
-(1.0327\pm0.0040)\partial_y[n_e\overline{ v_{ey}p_{ey}p_{ez}}]$\newline$-(1.0064\pm0.0044)n_e\overline{ p_{ey}} E_z$\newline$-(1.0099\pm0.0035)n_e\overline{ p_{ez}} E_y$\newline$-(1.0042\pm0.0027)n_e\overline{ v_{ex}p_{ey}} B_y$\newline $-(1.0061\pm0.0023)n_e\overline{ v_{ez}p_{ez}} B_x$\newline $+(1.0084\pm0.0027)n_e\overline{ v_{ex}p_{ez}} B_z$\newline $+(1.0031\pm0.0020)n_e\overline{ v_{ey}p_{ey}} B_x$ \\ \hline 
Boosted \newline Data \newline Inferred \newline Equation & $\partial_t [n_e\overline{ p_{ey}p_{ez}}]=$\newline$-(1.0016\pm0.0015)\partial_x[n_e\overline{ v_{ex}p_{ey}p_{ez}}]$\newline $
-(1.0008\pm0.0004)\partial_y[n_e\overline{ v_{ey}p_{ey}p_{ez}}]$\newline$-(0.9922\pm0.0033)n_e\overline{ p_{ey}} E_z$\newline$-(0.9912\pm0.0031)n_e\overline{ p_{ez}} E_y$\newline$-(0.9934\pm0.0029)n_e\overline{ v_{ex}p_{ey}} B_y$\newline $-(0.9908\pm0.0033)n_e\overline{ v_{ez}p_{ez}} B_x$\newline $+(0.9911\pm0.0031)n_e\overline{ v_{ex}p_{ez}} B_z$\newline $+(0.9921\pm0.0033)n_e\overline{ v_{ey}p_{ey}} B_x$ \\ \hline 

\end{tabular}
\end{center}
\begin{figure*}[t!]
    \centering
    \includegraphics[width=\textwidth]{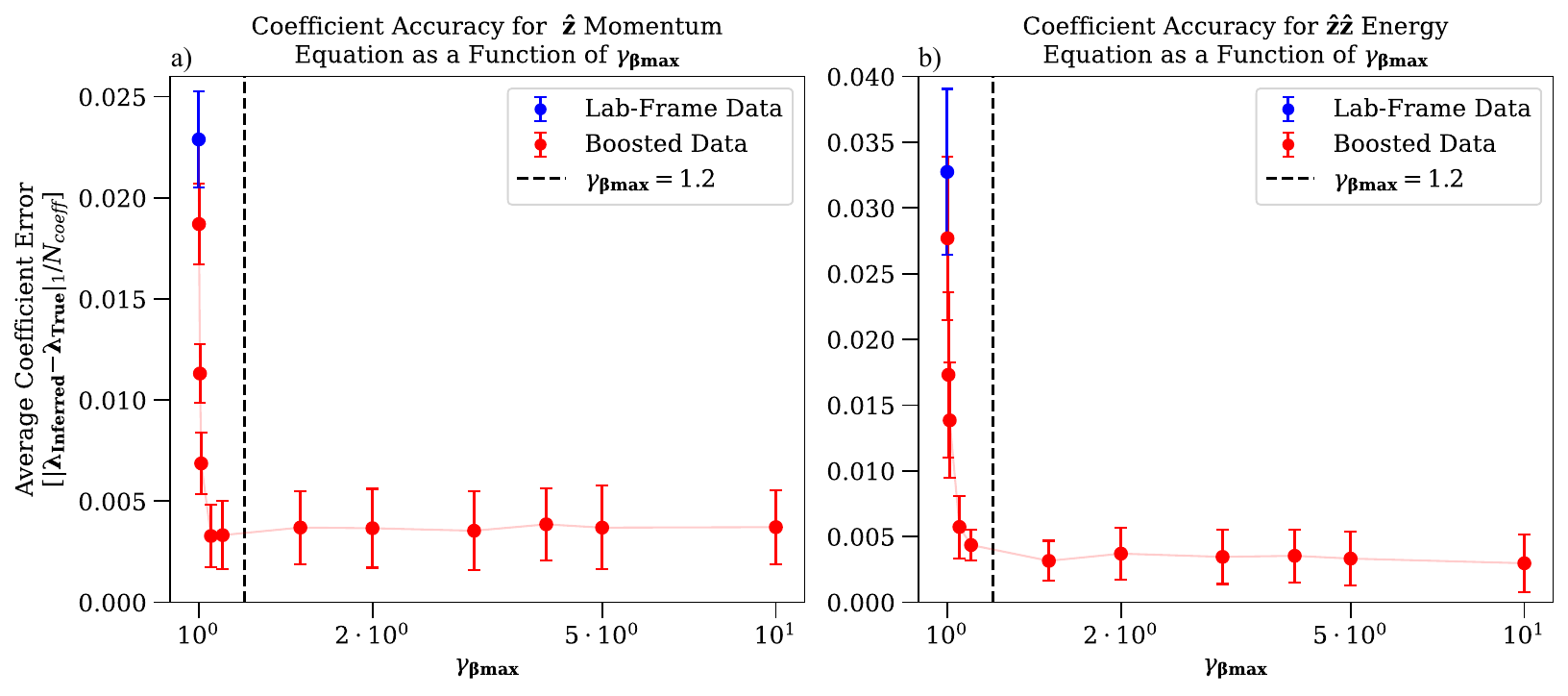}
    \caption{\textbf{Average coefficient error decreases as a function of maximum Lorentz boost ($\gamma_{\bm{\beta}max}$) for the $\bm{\hat{z}}$ momentum and $\bm{\hat{z}}\bm{\hat{z}}$ energy equations.} Plot a) is for the $\bm{\hat{z}}$ momentum equation and plot b) is for the $\bm{\hat{z}}\bm{\hat{z}}$ energy equation. The vertical axis measures the L1 difference between true and SR inferred coefficients. We randomly Lorentz transform our data in the range $1\leq \gamma_{\bm{\beta}}\leq\gamma_{\bm{\beta}max}$, and demonstrate that increasing $\gamma_{\bm{\beta}max}$ (embedding the symmetry more stringently) leads to more accurate coefficients. The vertical error bars correspond to $\pm$ one standard deviation in average coefficient error, and the statistics come from training each model five times with different realizations of the random data sampling.}
    \label{fig7}
\end{figure*}
\section{\label{app4}Additional Examples of Coefficient Error Improving as a Function of Maximum Boost}
Here we show two additional average coefficient error vs. $\gamma_{\bm{\beta }max}$ plots for the $\bm{\hat{z}}$ component of the momentum equation and $\bm{\hat{z}}\bm{\hat{z}}$ component of the energy equation [Figure \ref{fig7}]. Again the elbows fall roughly at $\gamma_{\bm{\beta }max}=1.2$. For the results presented in Figures \ref{fig1}, \ref{fig2}, \ref{fig4}, \ref{fig5}, and \ref{fig8} $\gamma_{\bm{\beta }max}=1.5$, to the right of the elbow. This demonstrates that the Lorentz boosts are large enough to reap the benefits of symmetry embedding. 

\section{\label{app6}Respecting Lorentz Invariance}

\begin{figure*}[t!]
    \centering
    \includegraphics[width=\textwidth]{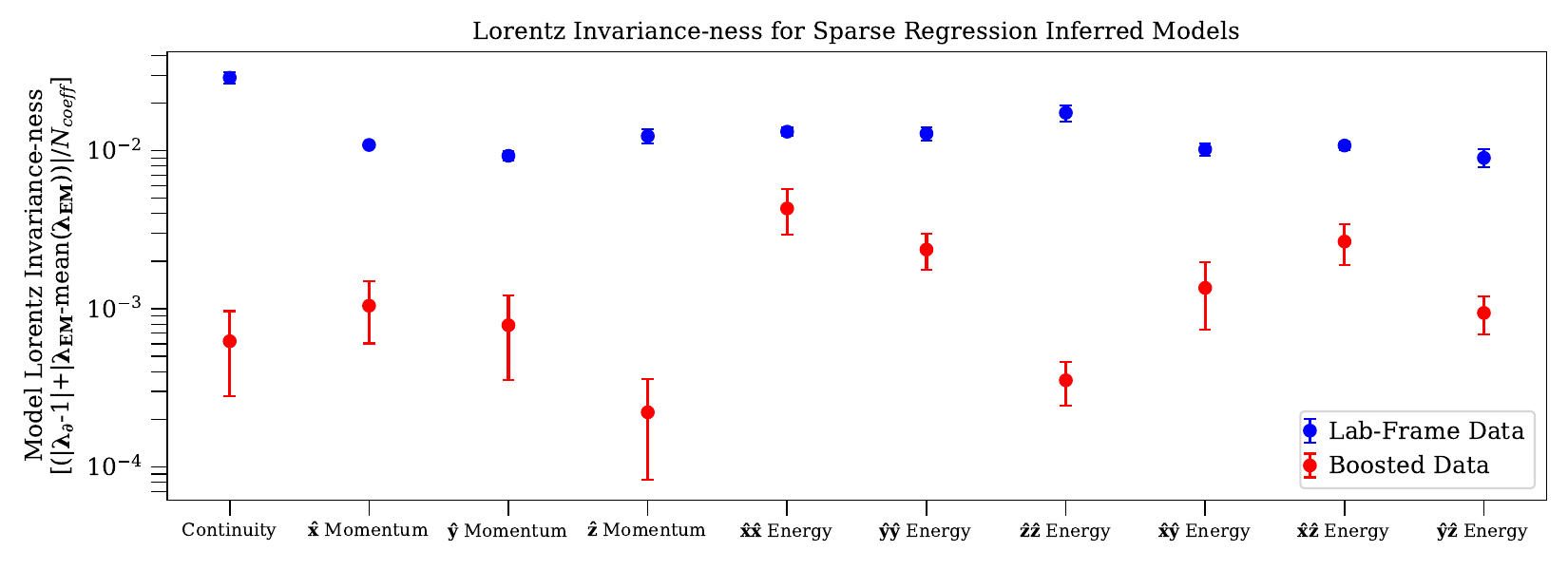}
    \caption{\textbf{Demonstrating that sparse regression inferred models with Lorentz-transformed data respect Lorentz invariance symmetry better than models discovered with only lab-frame data.} On the vertical axis we have constructed a metric to describe how well Lorentz invariance is satisfied by each sparse regression inferred model. A model which is perfectly Lorentz invariant has a metric measurement of 0, and the larger the metric the more the symmetry is broken. The metric we use is the absolute sum of the difference between the spatial derivative term coefficients and $1$, plus the absolute difference of the terms proportional to the electromagnetic fields and their mean. We normalize by the number of non-time derivative terms in the model. The vertical error bars correspond to $\pm$ one standard deviation, and the statistics come from training each model five times with different realizations of the random data sampling. On the horizontal axis are the first ten unique two-fluid plasma equations. We see that all ten models inferred with Lorentz-transformed data abide by Lorentz symmetry better than their lab-frame data counterparts beyond a standard deviation.}
    \label{fig8}
\end{figure*}
\begin{figure*}[t!]
    \centering
    \includegraphics[width=\textwidth]{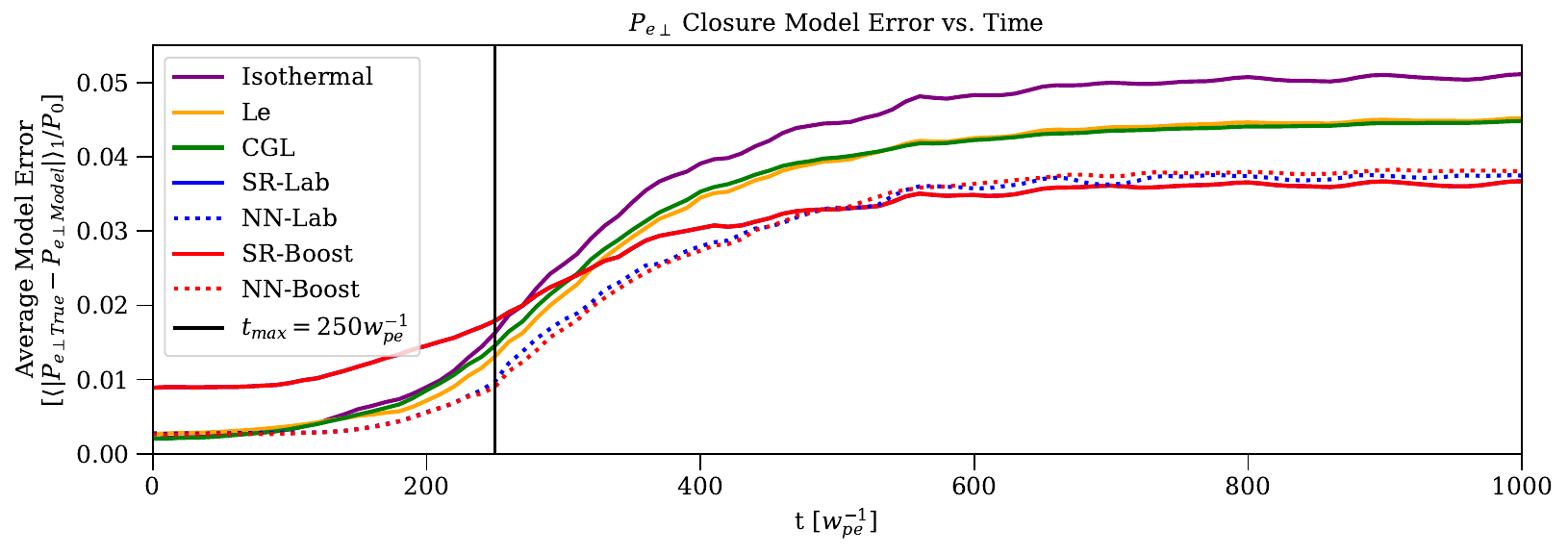}
    \caption{\textbf{Sparse regression and neural network $P_{e\perp}$ closures inferred from collisionless magnetic reconnection out-perform commonly used analytical closures, however no improvement from Galilean augmentation.} The vertical axis corresponds to model error measured as the spatial mean of the absolute L1 difference between the true simulation pressure and the pressure closure model. We normalize the average pressure error to a characteristic pressure value in the simulation $P_0=0.1 n_{0}m_ec^2$. The horizontal axis is time. The sparse regression and neural network models are trained until $t=t_{max}=250w_{pe}^{-1}$. As time increases and reconnection becomes more non-linear and complicated all closures perform worse. At late, generalization times the sparse regression and neural network models with and without boosting have the lowest model error. Since the sparse regression closures with and without Galilean data augmentation are identical, their error is identical. There is also very little difference between the neural network closures trained with and without data augmentation.}
    \label{fig11}
\end{figure*}
Reduced plasma models discovered with SR on Lorentz-augmented MR data are found to be more ``Lorentz invariant'' than models discovered on only the original, lab-frame simulation data. The symmetry is thus successfully embedded. Highly boosted Lorentz reference frames observe strikingly distorted dynamics relative to the simulation frame~\cite{Vay2009}; the balance of PDE terms in the boosted-frame data is different than in the lab-frame. By including Lorentz-augmented data in the equation inference procedure we are providing new, unseen, yet physically valid training data, which further restrict the solution space (coefficient manifold). 

We construct a metric to measure the degree of Lorentz invariance of our inferred two-fluid equations, and compare the equations discovered with Lorentz-boosted data vs. with lab-frame data. As a metric for the degree of Lorentz invariance from our inferred equations we add the absolute difference between the spatial derivative coefficients and $1$ to the absolute difference between the electromagnetic field term coefficients and their collective mean. In Figure \ref{fig8}, we show that all equations which are inferred with Lorentz-transformed data are more Lorentz invariant than when inferred solely with lab-frame data according to this metric.

To better understand how Lorentz boosting embeds Lorentz symmetry (and minimizes the Lorentz invariance metric described above), we return to the loss function formalism presented in Equations \ref{Eqn8}-\ref{Eqn11} and examine the regression loss function for the inference of the $\bm{\hat{z}}$ momentum equation with data that has been Lorentz transformed by $\bm{\beta}=\beta\bm{\hat{x}}$.
\begin{equation}\label{Eqn12}
    \mathcal{L}'\sim \partial_t'[n_e'\overline{ p_{ez}'}]+\lambda_1' n_e'E_z' + \lambda_2' n_e'\overline{ v_{ex}'} B_y' + \lambda_3' n_e'\overline{ v_{ey}'} B_x'
\end{equation}
\begin{figure*}[t!]
    \centering
    \includegraphics[width=\textwidth]{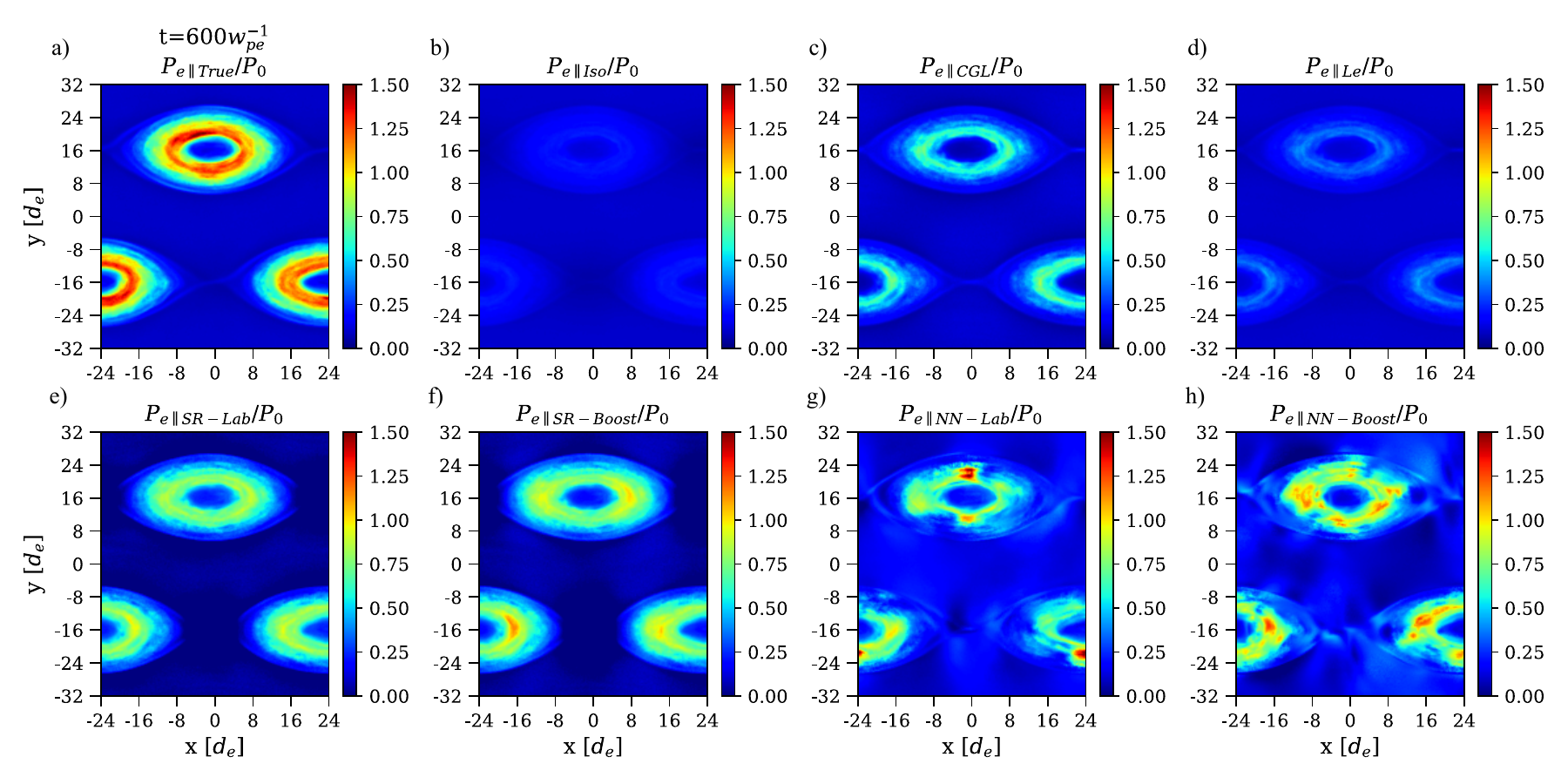}
    \caption{\textbf{Comparison of spatial patterns for analytical and ML inferred $P_{e\parallel}$ closure models.} In this plot we examine the spatial distributions of all closure models, evaluated on the PIC MR data in the $(x,y)$ plane at $t=600w_{pe}^{-1}$. This time is well into the generalization/testing phase. Panel (a) shows the true PIC pressure. Panel (b) depicts the Isothermal closure model evaluated on the PIC data, constructed with the PIC density. Likewise, panels (c) and (d) show the CGL and Le closure models respectively. The lower row is the ML inferred pressure closure models. Panels (e), (f), (g), and (h) plot the pressure as predicted by the SR-Lab, SR-Boost, NN-Lab, and NN-Boost models respectively. All pressure values are normalized to $P_0=0.1 n_{0}m_ec^2$, a characteristic downstream pressure value in the simulation.}
    \label{fig9}
\end{figure*}
\begin{figure*}[t!]
    \centering
    \includegraphics[width=\textwidth]{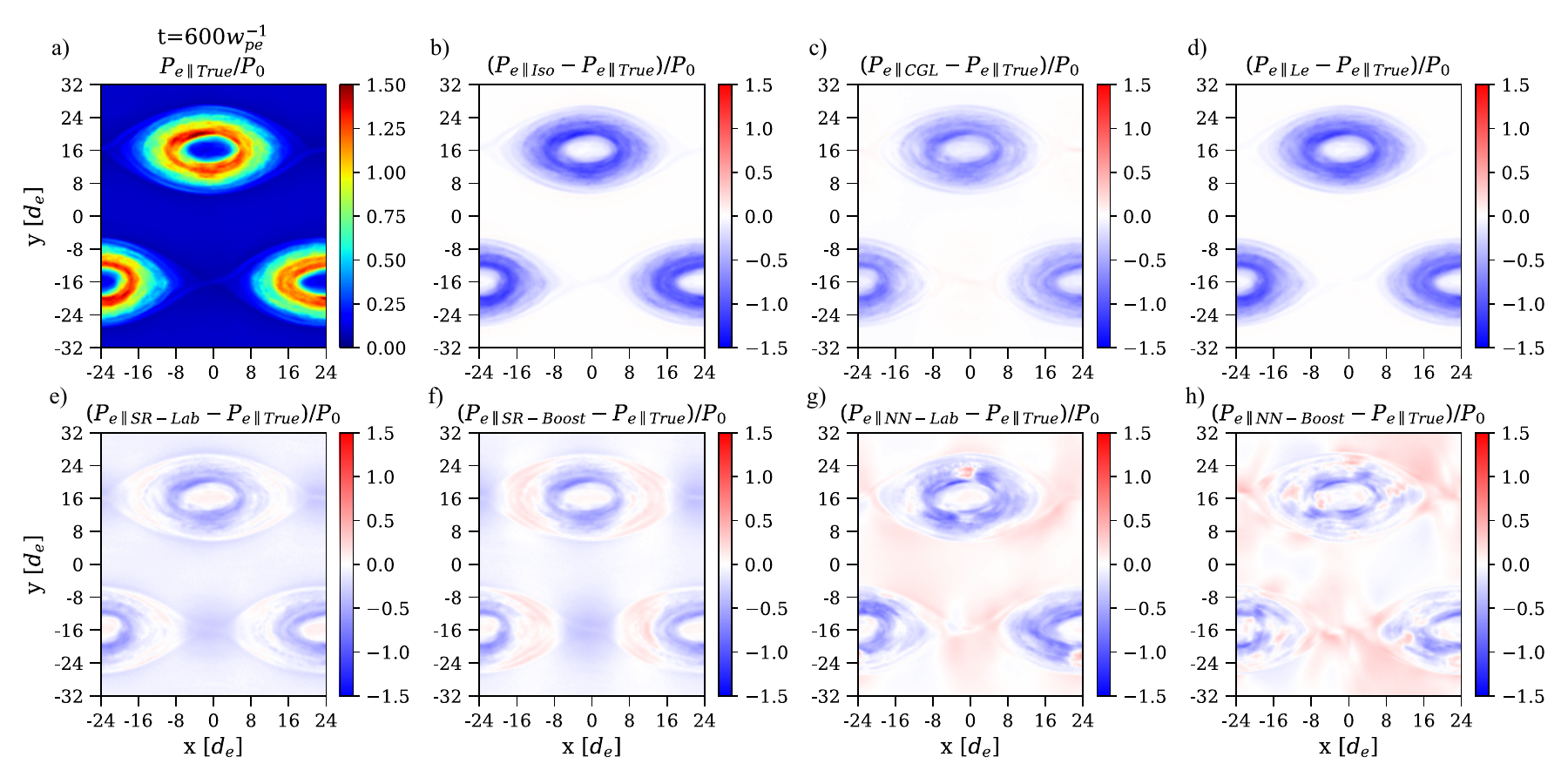}
    \caption{\textbf{Comparison of spatial errors for analytical and ML inferred $P_{e\parallel}$ closure models.} In this plot we examine the spatial distribution of error of all closure models, evaluated on the PIC MR data in the $(x,y)$ plane at $t=600w_{pe}^{-1}$. This time is well into the generalization/testing phase. Panel (a) shows the true PIC pressure, for reference. Panel (b) depicts the difference between the Isothermal closure model and the PIC pressure. Likewise, panels (c) and (d) plot the error of CGL and Le closure models respectively. The lower row is the error from the ML inferred pressure closure models. Panels (e), (f), (g), and (h) plot the errors of the SR-Lab, SR-Boost, NN-Lab, and NN-Boost models respectively. All pressure values are normalized to $P_0=0.1 n_{0}m_ec^2$, a characteristic downstream pressure value in the simulation. Positive (red) values indicate that the model pressure is larger than the true PIC pressure, while negative (blue) values indicate that the model pressure underestimates.}
    \label{fig10}
\end{figure*}

Where $\{\lambda_1',\lambda_2',\lambda_3'\}$ are unknown coefficients that the regression will try to optimize. From the Lorentz-boosted data we have measurements of the terms $\partial_t'[n_e'\overline{ p_{ez}'}]$, $n_e'E_{z}'$, etc. Using the inverse Lorentz transform $(\Lambda^{-1})^{\mu}_\nu$ we can express $\mathcal{L}'$ in terms of the lab-frame moments. Crucially we define the unknown coefficients to be invariant under the inverse Lorentz transformation. Focusing solely on the Lorentz force terms in Equation \ref{Eqn12}:
\begin{align}
    \mathcal{L}'\sim &\lambda_1 \gamma_\beta^2n_e[1-\beta\overline{ v_{ex}}][E_z+\beta B_y] 
    \nonumber \\ 
    &+\lambda_2 \gamma_\beta^2n_e[\overline{ v_{ex}}-\beta] [B_y+\beta E_z] +\lambda_3n_{e}\overline{ v_{ey}} B_x
\end{align}
Group like terms proportional to $n_eE_z$.
\begin{equation}
    \mathcal{L}'\sim n_eE_z \{ \lambda_1 [1-\beta\overline{ v_{ex}}]+\lambda_2[\beta\overline{ v_{ex}}-\beta^2]\}
\end{equation}
Take the limit $\gamma_{\bm{\beta}{max}}\rightarrow \infty $, corresponding to $\beta\rightarrow 1$.
\begin{equation}\label{Eqn15}
    \mathcal{L}'\sim [\lambda_1-\lambda_2][1-\overline{ v_{ex}}]n_eE_z
\end{equation}
Measurements of the term $[1-\overline{ v_{ex}}]n_eE_z$ from Equation \ref{Eqn15} are not typically equal to zero in the MR data set, thus to minimize the loss function in the large Lorentz boost limit the multiplicative term $[\lambda_1-\lambda_2]\rightarrow 0$, $\lambda_1\rightarrow \lambda_2$. Lorentz boosting forces the terms proportional to electromagnetic fields to converge to the same value. For the inferred equation to be Lorentz invariant the coefficients of terms proportional to electromagnetic field indeed must be identical in magnitude.

\section{\label{app7}Sparse Regression and Neural Network Perpendicular Pressure Closure Discovery for Magnetic Reconnection}
In Section \ref{sec:level2:2} we demonstrated the significant improvements when incorporating Galilean data augmentation in the sparse regression and neural network inference of a parallel pressure closure for collisionless pair-plasma magnetic reconnection. However, for a complete closure model we must also discover a pressure closure perpendicular to the local magnetic field. For example both CGL and Le closures have separate functional forms, descriptions, for the parallel and perpendicular pressure components. 

We repeated the sparse regression algorithm with lab-frame data, but now with our target variable as $P_{e\perp}$ rather than $P_{e\parallel}$. We allow $P_{e\perp}$ to depend on $\{n_e,\overline{v_{ex}},\overline{v_{ey}},\overline{v_{ez}},B_x,B_y,B_z\}$ and their second order polynomial combinations; again we do not include the electric field as a building block catalog term given its correlation with $\overline{\bm{v_e}}\times\bm{B}$ which crashes our sparse regression algorithm. As in the parallel pressure closure case we trained the sparse regression algorithm with $1,000$ data points sampled every $10w_{pe}^{-1}$ from $t=0w_{pe}^{-1}$ to $t=t_{max}=250w_{pe}^{-1}$, and only selected data points where $P_{e\perp}$ is greater than its $70$th percentile value at its sampled timestep. SR identified a five term electron perpendicular pressure model of the form:
\begin{align}\label{E1}
    P_{e\perp Lab} = & -0.004+0.021n_e+0.002n_e^2
    \nonumber 
    \\ & -0.152B_x^2-0.011B_zn_e
\end{align}
 We immediately notice that the SR inferred closure in Equation \ref{E1} is Galilean invariant, and thus there are no non-symmetry abiding terms to be removed like in the parallel pressure closure example. By performing Galilean data augmentation we double the training dataset size with random Galilean boosts of each sampled point with $|\beta_j|\leq 0.1c$ and inferred the same five-term electron perpendicular pressure model:

\begin{align} \label{E2}
    P_{e\perp Boost} = & -0.004+0.021n_e+0.002n_e^2
    \nonumber 
    \\ & -0.152B_x^2-0.011B_zn_e
\end{align}
 While incorporating Galilean-augmented data into sparse regression had no impact on the functional form of the perpendicular pressure closure, discovering the same model with Galilean-augmented data gives reassurance that the functional form is indeed Galilean invariant. We note that since the model trained with lab-frame data is already symmetry abiding, there is no amount of lab-frame data or severity of maximum boost velocity that could alter the original model. We see the sparse regression inferred closures out-performed commonly used analytical closures [See Figure \ref{fig11}]. 
Next, we trained neural networks to predict $P_{e\perp}$ with and without Galilean data augmentation to compare with the sparse regression model. We allowed $P_{e\perp}$ to depend on $\{n_e,\overline{v_{ex}},\overline{v_{ey}},\overline{v_{ez}},E_x,E_y,E_z,B_x,B_y,B_z\}$. Again, we sample quantities at 1,000 spatial locations every $10w_{pe}^{-1}$ from $t_{min}=0$ to $t_{max}=250w_{pe}^{-1}$. We only sample from spatial locations where $P_{e\parallel}$ is greater than its $70$th percentile value at its sampled timestep. When performing Galilean data augmentation we double the training
dataset size by including random Galilean boosts of each sampled point with $|\beta_j|\leq 0.1c$. Our network geometry is three hidden layers with 25 nodes per layer. Each network is trained in PyTorch for 2,000 ephochs with an AdamW optimizer, learning rate of 0.001, and batch size of 10. Additionally, each network is trained and tested five times to develop statistics for the performance of each model. We see that including Galilean-augmented data does not greatly improve model generalization [Figure \ref{fig11}]. Both neural network models outperform commonly used analytical pressure closures.
We see that the machine learning algorithms - both sparse regression and neural networks - did not rely on the non-Galilean invariant inputs to predict $P_{e\perp}$. Including Galilean-augmented data did not change model generalization. This is possibly because dynamics are more rapid and kinetic in the $\hat{\parallel}$ direction along magnetic field lines than in the $\hat{\perp}$ direction, thus the $P_{e\perp}$ evolution is much less interesting and easier to predict without the need for symmetry embedding to further restrict the solution space. We are operating in the medium guide field regime where particles are significantly tied to the field lines, but if we ventured into the low guide field limit the $P_{e\perp}$ dynamics would become more varied and perhaps symmetry embedding would become more important for perpendicular closure identification. 
\begin{figure*}[t!]
    \centering 
    \includegraphics[width=\textwidth]{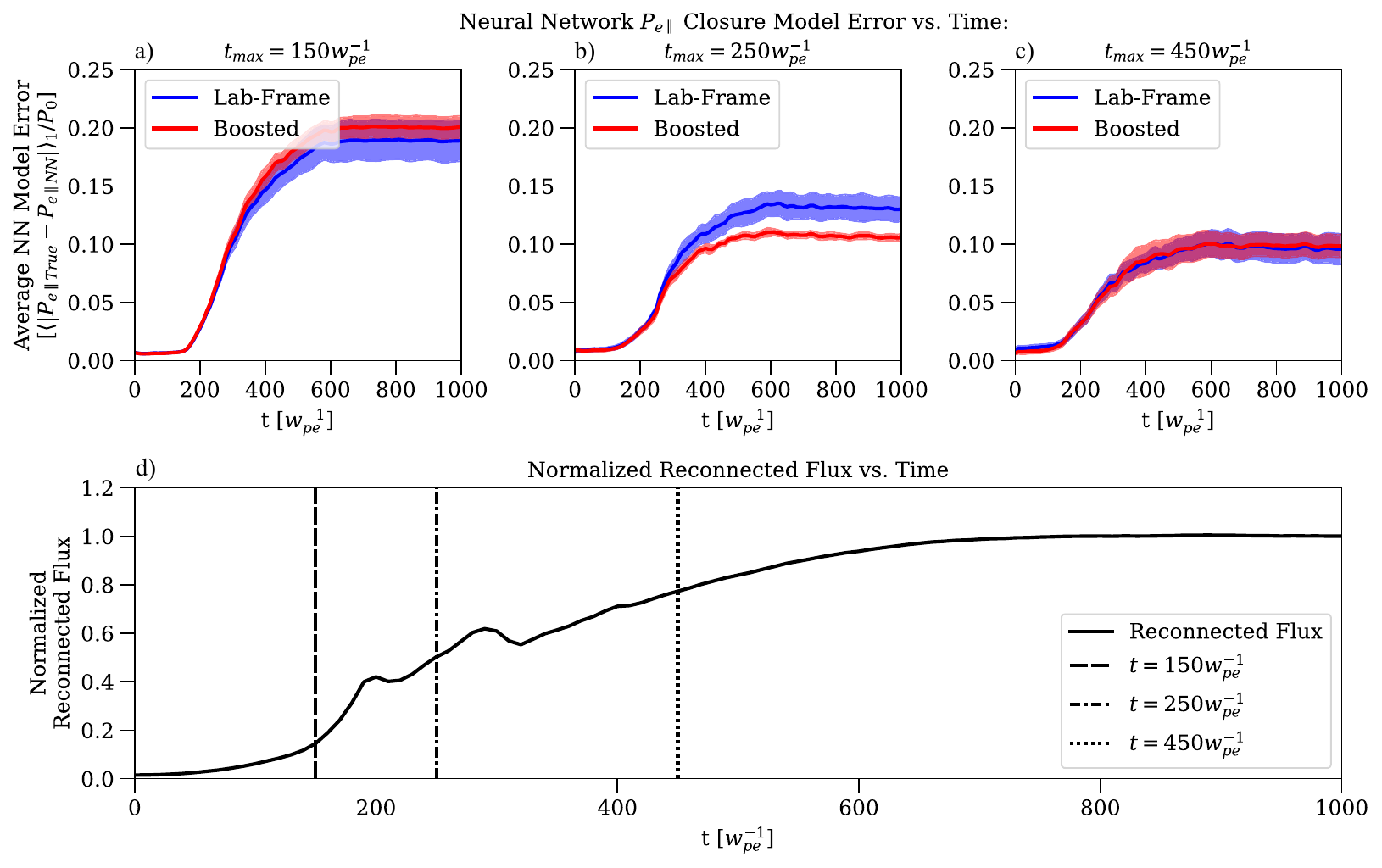}
    \caption{\textbf{Galilean data augmentation improves neural network closure model for $P_{e\parallel}$.} We train neural networks with and without Galilean-augmented data from $t=0w_{pe}^{-1}$ to differing $t_{max}$. $t_{max}$ is varied between $150w_{pe}^{-1}$, $250w_{pe}^{-1}$, and $450w_{pe}^{-1}$ across the top three plots a), b) and c) respectively. The vertical axes correspond to model error measured as the spatial mean of the absolute L1 difference between the true simulation pressure and the pressure closure model. The statistics come from training each model five times with different realizations of the random data sampling. We normalize the average pressure error to a characteristic pressure value in the simulation $P_0=0.1 n_{0}m_ec^2$. The horizontal axis is time.  In plot d) we display percentage of reconnected magnetic flux with the corresponding values of $t_{max}$ overlaid. We observe that for the models trained up to $t_{max}=250w_{pe}^{-1}$ in plot b) Galilean data augmentation improves generalization. For the other two models in plots a) and c) the performance differences between lab-frame and boosted neural networks are within the variance.} 
    \label{fig12}
\end{figure*}
\section{\label{app8}Analysis of Spatial Error Distribution of ML Closure Models for $P_{e\parallel}$}
We can evaluate the analytical and machine learned pressure closure models discovered in Section \ref{sec:level2:2} on the PIC MR data and examine the spatial distributions of their predictions [See Figure \ref{fig9}]. The pressure values are normalized to a characteristic downstream pressure value in the simulation $P_{0}=0.1 n_{0}m_ec^2$. We plot the spatial distributions at $t=600w_{pe}^{-1}$, significantly later than the ML training cutoff time $t_{max}=250w_{pe}^{-1}$. The true PIC simulation $P_{e\parallel}/P_0$ is plotted in Figure \ref{fig9} (a). We see that the analytical closure models - Isothermal [Figure \ref{fig9} (b)], CGL [Figure \ref{fig9} (c)], and Le [Figure \ref{fig9} (d)] - all greatly under predict the pressure inside of the plasmoids at this time. The ML closure models - SR-Lab [Figure \ref{fig9} (e)], SR-Boost [Figure \ref{fig9} (f)], NN-Lab [Figure \ref{fig9} (g)], and NN-Boost [Figure \ref{fig9} (h)] - predict larger, more accurate pressure values than the analytical models inside of the plasmoids. 

To better understand where in the spatial domain different models succeed or fail, we also visualize the differences between model predictions and the true PIC $P_{e\parallel}$ [See Figure \ref{fig10}]. For reference, the true PIC pressure is again shown in the upper left panel, Figure \ref{fig10} (a). From this error plot it becomes easier to see that the analytical models - Isothermal [Figure \ref{fig10} (b)], CGL [Figure \ref{fig10} (c)], and Le [Figure \ref{fig10} (d)] - greatly underestimate the pressure inside of the plasmoids, while correctly modeling the upstream regions outside of the plamoids. In contrast, the ML models - SR-Lab [Figure \ref{fig10} (e)], SR-Boost [Figure \ref{fig10} (f)], NN-Lab [Figure \ref{fig10} (g)], and NN-Boost [Figure \ref{fig10} (h)] - better capture the high pressure plasmoid regions, yet perform worse outside of the plasmoids. Such behavior is likely a result of our training data sampling method, which selects only those measurements where the target variable $P_{e\parallel}$ is greater than its $70$th percentile value at its sampled timestep. This data bias leads to an imbalance in the data set, causing the ML models to prioritize fitting larger pressure regions inside of the plasmoids and de-emphasizing smaller pressure regions outside of the plasmoids. Future work should explore the impact of more balanced data sets.

The main improvement of SR-Boost compared with SR-Lab is its improved accuracy in predicting regions outside of the plasmoids [Figure \ref{fig10}], which leads to the lower mean absolute error [Figure \ref{fig6}]. As for NN-Boost over NN-Lab, the predominant improvement is inside of the plasmoids. Lastly, it is interesting to note that the SR-Boost and NN-Boost models have similar mean absolute errors at $t=600w_{pe}^{-1}$, yet their spatial error patterns have different structures. This indicates that SR and NN models have identified different functional forms for the closure, yet the NN's highly non-linear expressive functional form is superfluous as it does not improve mean absolute closure error relative to SR in this phase.

\section{\label{app9}Impact of Maximum Training Time on the Relative Performance of Lab-Frame and Boosted Neural Networks}
In Section \ref{sec:level2:2} we discovered $P_{e\parallel}$ closures with data $t<t_{max}=250w_{pe}^{-1}$ (corresponding to a time when $\sim50\%$ of the magnetic flux has reconnected) and then examined pressure closure model generalization to later times. $t_{max}$ is a free parameter which we can vary, and here we examine neural network lab-frame and boosted models trained with different values of $t_{max}$. We vary $t_{max}=\{150,250,450\}w_{pe}^{-1}$ to understand when Galilean data augmentation improves model generalization and when it does not [Figure \ref{fig12}]. We additionally train each model five separate times with different realizations of the random data sampling to ensure our results are statistically significant given the spread in model performance for different samplings. We find that for the models trained with data up to $t_{max}=150w_{pe}^{-1}$, the network is not able to improve with the addition of Galilean boosted data [Figure \ref{fig12} (a)]. We believe this is due to the lack of sufficiently rich (nonlinear) dynamics in the training data, which prevents the ML model from effectively capturing the underlying physics. As a result, Galilean-augmented data does not lead to improved model performance. For the models trained with data up to $t_{max}=250w_{pe}^{-1}$, Galilean-augmented data gives a significant generalization improvement [Figure \ref{fig12} (b)]. Lastly, for the models trained with data up to $t_{max}=450w_{pe}^{-1}$, we again observe no improvement with the inclusion of Galilean-augmented data [Figure \ref{fig12} (c)]. This is presumably  because the dynamics at times greater than $t_{max}=450w_{pe}^{-1}$ do not significantly differ than those during the training data, hence the model trained with lab-frame data is not greatly penalized by not abiding by the symmetry and grasping onto non-symmetry preserving terms. There appears to be a Goldilocks zone ($t=250w_{pe}^{-1}$) in which the models have been exposed to enough, but too much, dynamics allowing Galilean-augmented data to meaningfully improve generalization.

%\nocite{*}

\bibliography{mybib}% Produces the bibliography via BibTeX.

\end{document}